\documentclass[aps,preprint,superscriptaddress,showpacs,preprintnumbers,showkeys,amsmath,amssymb]{revtex4}
\usepackage[dvips]{graphicx}
\usepackage{dcolumn}
\usepackage{amsmath}
\usepackage{amssymb}
\usepackage{amsbsy}
\usepackage{units}
\usepackage{float}

\usepackage{bm}

\usepackage{color}
\usepackage{ulem}

\begin{document}
\title{A comprehensive study of the magnetic, structural and transport properties of the III-V ferromagnetic semiconductor InMnP}
\author{M. Khalid}
\affiliation{%
Helmholtz-Zentrum Dresden Rossendorf, Institute of Ion Beam Physics and Materials Research, Bautzner Landstrasse 400, D-01328 Dresden, Germany}%
\author{Kun Gao}
\affiliation{%
Helmholtz-Zentrum Dresden Rossendorf, Institute of Ion Beam Physics and Materials Research, Bautzner Landstrasse 400, D-01328 Dresden, Germany}%
\affiliation{%
Technische Universit$\ddot{a}$t Dresden, 01062 Dresden, Germany}%
\author{E. Weschke}
\affiliation{%
Helmholtz-Zentrum Berlin f$\ddot{u}$r Materialien und Energie, Wilhelm-Conrad-R$\ddot{o}$ntgen-Campus BESSY II, D-12489 Berlin, Germany}%
\author{R. H\"ubner}
\affiliation{%
Helmholtz-Zentrum Dresden Rossendorf, Institute of Ion Beam Physics and Materials Research, Bautzner Landstrasse 400, D-01328 Dresden, Germany}%
\author{C. Baehtz}
\affiliation{%
Helmholtz-Zentrum Dresden Rossendorf, Institute of Ion Beam Physics and Materials Research, Bautzner Landstrasse 400, D-01328 Dresden, Germany}%
\author{O. Gordan}
\affiliation{%
Institute of Physics, Chemnitz University of Technology, 09107 Chemnitz, Germany}%
\author{G. Salvan}
\affiliation{%
Institute of Physics, Chemnitz University of Technology, 09107 Chemnitz, Germany}%
\author{D.R.T. Zahn}
\affiliation{%
Institute of Physics, Chemnitz University of Technology, 09107 Chemnitz, Germany}%
\author{W. Skorupa}
\affiliation{%
Helmholtz-Zentrum Dresden Rossendorf, Institute of Ion Beam Physics and Materials Research, Bautzner Landstrasse 400, D-01328 Dresden, Germany}%
\author{M. Helm}
\affiliation{%
Helmholtz-Zentrum Dresden Rossendorf, Institute of Ion Beam Physics and Materials Research, Bautzner Landstrasse 400, D-01328 Dresden, Germany}%
\affiliation{%
Technische Universit$\ddot{a}$t Dresden, 01062 Dresden, Germany}%
\author{Shengqiang Zhou}
\email[Electronic address: ]{S.Zhou@hzdr.de}
\affiliation{%
Helmholtz-Zentrum Dresden Rossendorf, Institute of Ion Beam Physics and Materials Research, Bautzner Landstrasse 400, D-01328 Dresden, Germany}%
\date{\today}
\begin{abstract}
The manganese induced magnetic, electrical and structural modification in InMnP epilayers, prepared by Mn ion implantation and pulsed laser annealing, are investigated in the following work. All samples exhibit clear hysteresis loops and strong spin polarization at the Fermi level. The degree of magnetization, the Curie temperature and the spin polarization depend on the Mn concentration. The bright-field transmission electron micrographs show that InP samples become almost amorphous after Mn implantation but recrystallize after pulsed laser annealing. We did not observe an insulator-metal transition in InMnP up to a Mn concentration of 5 at.$\%$. Instead all InMnP samples show insulating characteristics up to the lowest measured temperature. Magneotresistance results obtained at low temperatures support the hopping conduction mechanism in InMnP. We find that the Mn impurity band remains detached from the valence band in InMnP up to 5 at.$\%$ Mn doping. Our findings indicate that the local environment of Mn ions in InP is similar to GaMnAs, GaMnP and InMnAs, however, the electrical properties of these Mn implanted III-V compounds are different. This is one of the consequences of the different Mn binding energy in these compounds.    
\end{abstract}
\maketitle
\section{Introduction} 
After more than two decades, the study of dilute magnetic semiconductors has developed into an important branch of materials science. The comprehensive investigation of dilute magnetic III-V semiconductors has been stimulated by successful demonstrations of several phenomenological functionalities in these types of materials. For instance, III-V:Mn semiconductors exhibit properties like spin injection \cite{Ohno790} and the control of magnetism by means of an electric field \cite{Ohno944,Chiba515}. It has been demonstrated (with anomalous Hall signals) that the ferromagnetism in III-V semiconductor is carrier-mediated \cite{Scarpulla207204}. These properties of III-V:Mn semiconductors make them highly suitable for spintronic device applications \cite{Wolf1488}.\newline
Despite several outstanding achievements, the origin and control of ferromagnetism in III-V semiconductors is one of the most controversial research topics in condensed-matter physics today. GaMnAs is the most studied and well understood III-V dilute magnetic semiconductor. Currently there are two main competing theories under discussion for explaining the ferromagnetism in III-V semiconductors, particularly in GaMnAs. The first one states that a strong hybridization of Mn 3$d$-electrons with the GaAs valence band occurs. As a result, the Mn-derived band (states) merges with the GaAs valence band, giving rise to hole-mediated ferromagnetism through the $p-d$ exchange interaction \cite{Dietl1019,Masek227202}. The second type states that the Mn states are split off from the valence band and they lie in an impurity band in the bandgap about 110 meV above the valence band maximum. In this scenario, the ferromagnetism is explained by the double-exhange interaction \cite{Burch087208,Hirakawa193312,Ando067204,Rokhinson161201,Ohya342}.\newline
The Mn concentration and the hole concentration are the key factors in controlling the Curie temperature and the strength of the exchange interaction in III-V semiconductors, e.g. GaMnAs, GaMnP \cite{Dobrowolska444,Scarpulla207204}. Mn-doped GaAs exhibits insulator-metal transition at a certain Mn concentration \cite{Dobrowolska444}. This is due to the isolated acceptor states of the Mn ions in GaAs being only 110 meV above the valence band maximum. With the introduction of the Mn ions, an impurity band forms and with increasing Mn concentration the band broadens, which results in a valence band like conduction \cite{Clerjaud3615}. In the case of GaMnP, the Mn isolated acceptor state is around 440 meV above the valence band, four times greater than that of GaMnAs, therefore the band broadening and the thermal energy might not be sufficient to induce metal behaviour in GaMnP even up to a high Mn concentration \cite{Scarpulla207204,yuan2401304}. Recently, we have shown that ferromagnetic order can be induced in InMnP with a Mn concentration which is comparable to that of GaMnAs and GaMnP \cite{Khalid121301}. The aim is to study a system that has the Mn acceptor level in between GaMnAs and GaMnP compounds in order to shed light on the impurity versus valence band debate in III-V semiconductors. InMnP is the most favourable choice to study this effect as it has a bandgap of 1.34 eV and an isolated Mn energy level of 220 meV \cite{Clerjaud3615}. \newline
In this work, we show how the variation in Mn concentration in InMnP modifies its magnetic, transport and structural properties. This is the first time that such a detailed and systematic study of the role of Mn in InMnP is carried out. This work contributes to a comprehensive understanding of impurity versus valence band picture in Mn-doped III-V semiconductors.     
\section{Experimental}
Semi-insulating InP (100) wafers were implanted with 50 keV Mn$^+$ ions with fluences of 5$\times$10$^{15}$/cm$^2$, 1$\times$10$^{16}$/cm$^2$, 1.5$\times$10$^{16}$/cm$^2$ and 2$\times$10$^{16}$/cm$^2$ at room temperature. Hereafter samples are named as A, B, C and D, respectively. The Mn-ion concentration and penetration depth in InP were estimated by SRIM (the stopping and range of ions in matter) \cite{Ziegler1027}. The implantation energy of Mn$^+$ ions was chosen in such a way that the penetration depth remains near 100 nm in InMnP. After implantation the samples were annealed by a XeCl excimer laser using an energy density of 0.40$\pm$0.05 J/cm$^2$ for a single pulse duration (30 ns). It is well known that after annealing a Mn-rich surface is created \cite{Scarpulla207204} which should be removed for further sample characterization. We have used a (1:10) HCl solution to etch the Mn-rich top surface from InMnP samples. The Mn-concentration measured by Auger electron spectroscopy (not shown here) in sample D was approximately 5 at.$\%$. The measured (estimated) Mn concentrations in samples A, B, C, and D are 1, 2, 3 and 5 at.$\%$, respectively. A SQUID-VSM (Superconducting Quantum Interference Device-Vibrating Sample Magnetometer) was used to measure the magnetization of the InMnP samples while magnetotransport measurements were performed using a Lakeshore system. X-ray Absorption Spectroscopy (XAS)/X-ray Magnetic Circular Dichroism (XMCD) measurements were performed at the beamline UE46/PGM-1 at BESSY II (Helmholtz-Zentrum Berlin). High resolution x-ray diffraction measurements were performed at the European synchrotron radiation facility (ESRF-BM20), in Grenoble, France. To locally analyze the microstructure of the Mn-implanted InP, transmission electron microscopy (TEM) investigations were performed using an image-corrected Titan 80-300 microscope (FEI). Besides bright-field imaging, selected area electron diffraction (SAED) was used to obtain structural information. Since the smallest available selected area aperture of 10 $\mu$m covers a circular area with a diameter of about 190 nm, both the Mn-implanted surface layer as well as the InP substrate contribute to the SAED patterns. Prior to each TEM analysis, the specimen mounted in a double tilt analytical holder was placed for about 30 s into a Model 1020 Plasma Cleaner (Fischione) to remove organic contamination. Classical cross-sectional TEM specimens were prepared by sawing, grinding, dimpling, and final Ar ion milling. Ultra-Violet Raman measurements were performed in the backscattering geometry using a 325 nm line of a He-Cd laser in a $\bar{z}$($\acute{y}$ , $\acute{y}$)z back-scattering configuration.
\section{Results and discussion}
In this section, the magnetic, structural and transport properties of four InMnP samples will be discussed in detail. An overview of the samples including their Mn concentration, Curie temperature, magnetization, activation energies, strain, magnetoresistance, value of the constant $C$ (see Eq.~\ref{2}) and XMCD signals is given in table~\ref{table1}.
\begin{center}
\begin{table*}[htb]
\caption{Magnetic, structural and transport properties of InMnP in this study.}
\label{table1}
\begin{tabular}{cccccccccc}
\hline
Material & Mn & T$_c$ & Magnetization & E$_1$ & E$_2$ & Strain & MR & $C$ & XMCD \\
 InMnP& (at.$\%$) & (K) & (emu/cm$^{-3}$) & (meV) & (meV) & ($\%$) & ($\%$, 5K and 50K) & - & ($\%$) \\
\hline
A & 1 & 10 & 2 & 57 & 8 & -0.33 & - and 20 & - & 5 \\
B & 2 & 15 & 4 & 35 & 6 & - & 59 and 22 & -12910 & 10 \\
C & 3 & 20 & 6 & 33 & 5 & -0.34 & 63 and 34 & -16321 & - \\
D & 5 & 40 & 10 & 20 & 4 & -0.35 & 65 and 39 & -20465 & 40 \\
\hline
\end{tabular}
\end{table*}
\end{center}
\subsection{Structural properties}
\begin{figure}[t]
\includegraphics[width=0.45\textwidth]{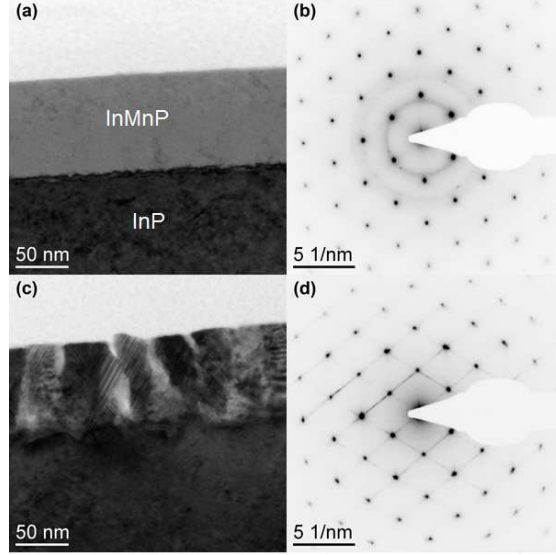}
\caption{\label{Fig.1} Cross-sectional bright-field TEM micrographs and corresponding selected area electron diffraction patterns of Mn-implanted InP before (a, b) and after laser annealing (c, d). The SAED patterns include diffraction information from both the implanted surface layer and the InP substrate.}
\end{figure}

The microstructure of the InMnP sample D with a Mn concentration of 5 at.$\%$ was investigated by transmission electron microscopy (TEM). Figure~\ref{Fig.1}(a) shows a cross-sectional bright-field TEM image of the as-implanted sample. The gray color of the approximately 90 nm thick surface layer on the single crystalline InP substrate indicates InP amorphization due to Mn implantation. However, the remaining diffraction contrast points to crystalline inclusions within the amorphous layer which were confirmed by selected area electron diffraction (SAED). In particular, Figure~\ref{Fig.1}(b) presents a SAED pattern with diffraction information from both, the InP substrate and the surface layer. Since the diffraction rings with uniformly distributed intensity are crossing the spots, which are caused by the single crystalline InP substrate, the inclusions in the surface layer are randomly oriented InP nanocrystallites. Laser annealing of the as-implanted sample changes the microstructure of the surface layer, as can be seen in the bright-field TEM micrograph in Figure~\ref{Fig.1}(c). A SAED pattern including diffraction information from the layer and the substrate indicates a quasi-epitaxial InP regrowth during laser annealing. However, stacking faults are introduced during this growth process, which can be inferred from the TEM micrograph in Figure~\ref{Fig.1}(c) and the lines in the SAED pattern of Figure~\ref{Fig.1}(d). Hence, the crystalline quality of the InMnP epilayer is comparable to that of a laser-annealed GaMnP epilayer \cite{Dissertation145}.\newline

\begin{figure}[t]
\begin{center}
\includegraphics[width=0.45\textwidth]{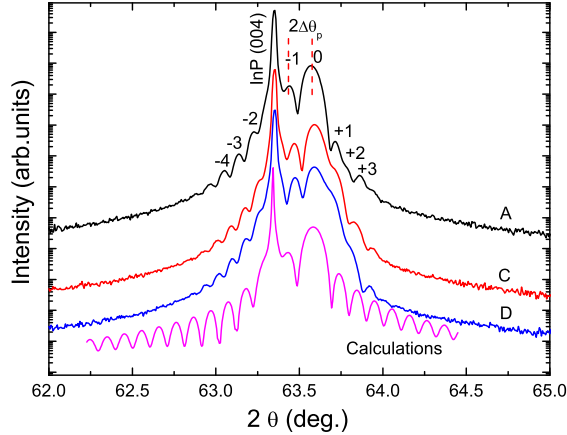}
\caption{\label{Fig.2} XRD pattern for three laser annealed InMnP samples A, C and D. A calculated XRD pattern of InMnP is also shown. Note that the x-axis is recalculated for $\lambda = 1.5405~{\textup{\AA}}$.}
\end{center}
\end{figure}
Figure~\ref{Fig.2} shows XRD scans taken from three InMnP samples A, C and D. The experiments are performed at the BM20 (ROBL) beamline at the European Synchrotron Radiation Facility (ESRF) Grenoble, France. The XRD scans include two peaks, a strong one which belongs to the InP substrate (004) reflection and a broad one which appears at a higher angle and belongs to the thin InMnP layer. Along with substrate (004) and InMnP (004) peaks, the presence of pendell\"osung fringes prove the reasonably good crystalline quality and interface. The InMnP (004) peak shifts to the higher angle with increasing Mn concentration which also means that the lattice constant decreases with increasing Mn concentration. As the lattice constant of the InMnP epilayer is smaller than that of the InP substrate, therefore, InMnP epilayers are under tensile strain. The XRD pattern is calculated theoretically and is shown in Fig.~\ref{Fig.2}. The calculation yields features similar to experimental ones. The tensile strain calculated for sample D is about $-0.35\pm 0.05$ $\%$. The tensile strain in the InMnP epilayer depends slightly on the Mn concentration.\newline
Furthermore, the angle between two satellite peaks (2$\Delta\theta_p$, as shown in Fig.~\ref{Fig.2}) is used to calculate the thickness of the InMnP epilayer, as this method has already been used for other layered crystalline systems \cite{Wu920}. The thickness of the InMnP epilayer is calculated by using the relationship L = $\lambda$/ (2$\Delta \theta_P$.cos$\theta_B$), where 2$\Delta\theta_P$ and $\theta_B$ are the angle between two satellite peaks and the Bragg angle, respectively. The thickness of sample D using this formula is found to be 95$\pm$5 nm, in agreement with SRIM simulation (not shown here) and TEM (see Fig.~\ref{Fig.1}(a)) results. Note that a shift in the InMnP peak with the Mn concentration is not as significant as in GaMnAs \cite{zhao2005intrinsic}.
Note also that the Mn interstitials and the In/Ga vacancies can induce the expansion of the lattice constant \cite{zhao2005intrinsic}. This expansion can compensate the shrinking induced by the substituted Mn in InP.\newline
\begin{figure}[t]
\begin{center}
\includegraphics[width=0.45\textwidth]{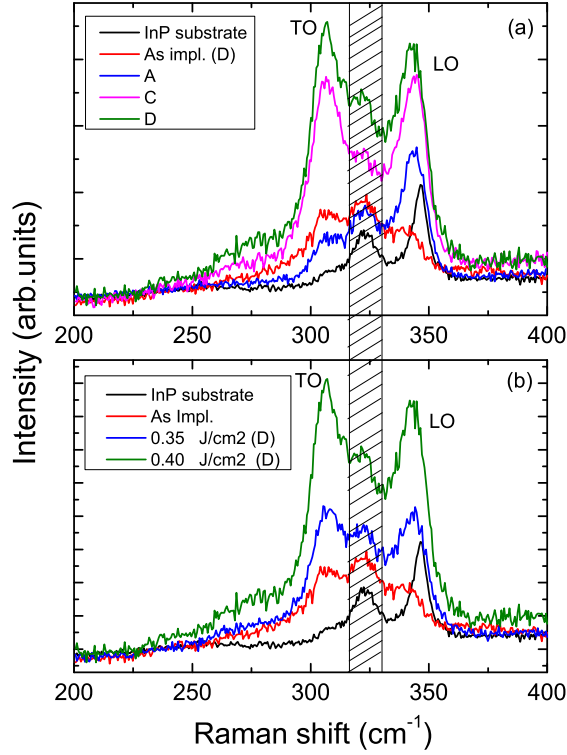}
\caption{\label{Fig.3} (a) Room temperature UV-Raman spectra of (100) oriented InMnP epilayers with different Mn concentrations along with a virgin InP substrate and an as-implanted InMnP sample. (b) Similar UV-Raman spectra of sample D annealed at two different energy densities along with a virgin and an as-implanted reference samples. The bumps (modes) marked with vertical shaded area are artifacts related to the experimental setup.}
\end{center}
\end{figure}
Figure~\ref{Fig.3} shows room temperature UV-Raman spectra of the samples A, C and D along with spectra of a virgin and an as-implanted samples. The UV-Raman spectrum of InP substrate consists of two peaks. The peak around 346 cm$^{-1}$ is due to InP longitudial phonon mode (LO) while the transverse-optical (TO) phonon mode for the InP (100) surface is forbidden according to the Raman selection rules \cite{Yu375,Trallero4030,Seong033202}. A broad peak or a shoulder at 323 cm$^{-1}$ originate from the experimental setup (glass, membrane for UV Light and marked with shaded area ). It is known that the longitudinal vibration of the carrier plasma is strongly damped due to the interaction between carrier plasmon and LO phonon via their macroscopic electric field, also known as a coupled plasmon-LO phonon (CPLP) mode. It has been demonstrated that the LO phonon mode is shifted to the position of TO phonon mode in GaMnAs only for a Mn concentration of 2.8 at.$\%$ \cite{Limmer205209}. It also indicates a high degree of incorporation of Mn in the GaAs lattice which ensures a high concentration of free carriers (holes) in this system.\newline
From Fig.~\ref{Fig.3}(a), it is clear that the virgin sample mainly exhibits a LO phonon mode which nearly vanishes after the Mn implantation due to the amorphization of the implanted layer. At the same time the normally forbidden TO phonon becomes visible, due to the relaxation of the selection rules caused by the Mn implantation. After laser annealing, sample A with a lowest Mn concentration exhibits a strong LO phonon mode at 346 cm$^{-1}$ along with a broad TO phonon mode at 307 cm$^{-1}$. As the Mn concentration increases, e.g., samples C and D, the intensity of the TO phonon mode at 307 cm$^{-1}$ increases and it is slightly higher than the LO phonon mode for the sample D. The increase in intensity of InP TO phonon upon increasing Ar irradiation doses was observed before and was attributed to an increase crystal lattice distortion \cite{Maslar2983}. It should be mentioned that even for Mn concentration of 5 at.$\%$  we do not observe the specific spectrum of amorphous InP in the annealed samples, which is a proof of the good crystallinity. Moreover, we do not observe the occurence of the coupled plasmon-LO mode which was reported in the case of GaMnAs. This indicates that the Mn ions incorporated in the crystal lattice do not contribute to a significant increase in free carriers density. This observation is in agreement with the high insulating behaviour and the hopping transport mechanism discussed in the section C.\newline
The incorporation of Mn ions into the GaAs lattice is more effective than in the InP lattice. As a result, these two compounds have different electronic and magnetic characteristics such as T$_c$, magnetization, carrier concentration, mobility etc. Furthermore, the coupled plasmon-LO phonon mode strongly depends on the free carrier mobility and concentration in III-V semiconductors \cite{Yuasa3962,Limmer205209}. Therefore, we do not expect identical behaviour of the LO phonon mode in GaMnAs and InMnP. Nevertheless, the UV-Raman results show that the InMnP epilayer recrystallizes after pulsed laser annealing and the LO, TO phonon intensity depends on the Mn concentration.
\subsection{Magnetic properties}
\begin{figure}[t]
\begin{center}
\includegraphics[width=0.45\textwidth]{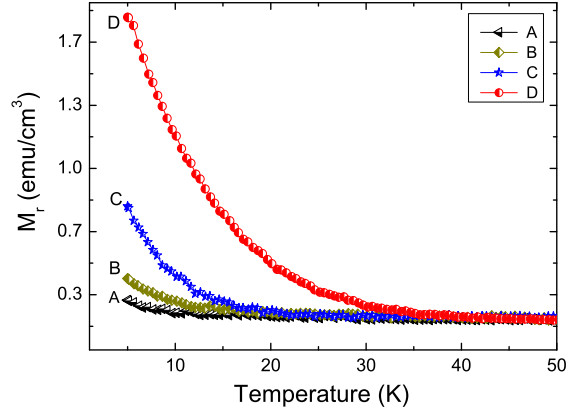}
\caption{\label{Fig.4} Remanent magnetization as a function of temperature for several laser annealed InMnP samples (A, B, C and D). The samples were first cooled down to 5 K under a 1 T magnetic field then the field was reduced to 20 Oe and the magnetizations were measured between 5 and 100 K for all samples. However, the $T_c$ obtained in this way might be overestimated since the possible superparamagnetic component in heavily compensated samples cannot be excluded \cite{Sawichi2010}. }
\end{center}
\end{figure}

Figure~\ref{Fig.4} shows the remanent magnetization as a function of temperature for four laser annealed InMnP samples A, B, C and D. The Mn concentration in these samples varies from approximately 1 to 5 at.$\%$. The Curie temperatures ($T_c$) of samples A, B, C and D are 10, 15, 20 and 40 K, respectively. The Curie temperature depends on both the Mn concentration (see Fig.~\ref{Fig.4}) and the annealing parameters (not shown here). We found that an annealing energy density of 0.4$\pm$0.05 J/cm$^2$ was the most suitable value for obtaining improved crystal quality, magnetic and transport properties of InMnP. A particular issue to be discussed here is the shape of M-T curves in InMnP samples. Most of the InMnP samples investigated in this study do not show the usual mean-field-like (Brillouin) behavior of spontaneous magnetization as a function of temperature. As the temperature increases from 5 K, there is a sharp decrease in the magnetization which finally slows down near the Curie temperature. This temperature dependence of magnetization in InMnP represents a concave shape of the M-T curve. Similar M-T curves have been observed in other III-V:Mn semiconductor compounds \cite{Ohno2664,Song2386} and have also been predicted theoretically \cite{Das155201}. Das Sarma $\it {et~al.}$ have theoretically shown that the concavity of the M-T curve is related to the ratio of the carrier concentration to the impurity concentration \cite{Das155201}. If the carrier/impurity ratio is low, the shape of the M-T curve will be more concave and vice versa. We expect that Mn ions replace indium ions at their lattice sites and generate holes in the InMnP system. But due to compensation centers, interstitial Mn or antisites as in GaMnAs \cite{edmonds2004mn, Lima012406}, which generally contribute electrons to the system, the free hole concentration in InMnP will be reduced. Consequently, the carrier(holes)/impurity ratio reduces significantly and hence a concave shaped M-T curve is observed in most of the InMnP samples.\newline 
\begin{figure}[t]
\begin{center}
\includegraphics[width=0.45\textwidth]{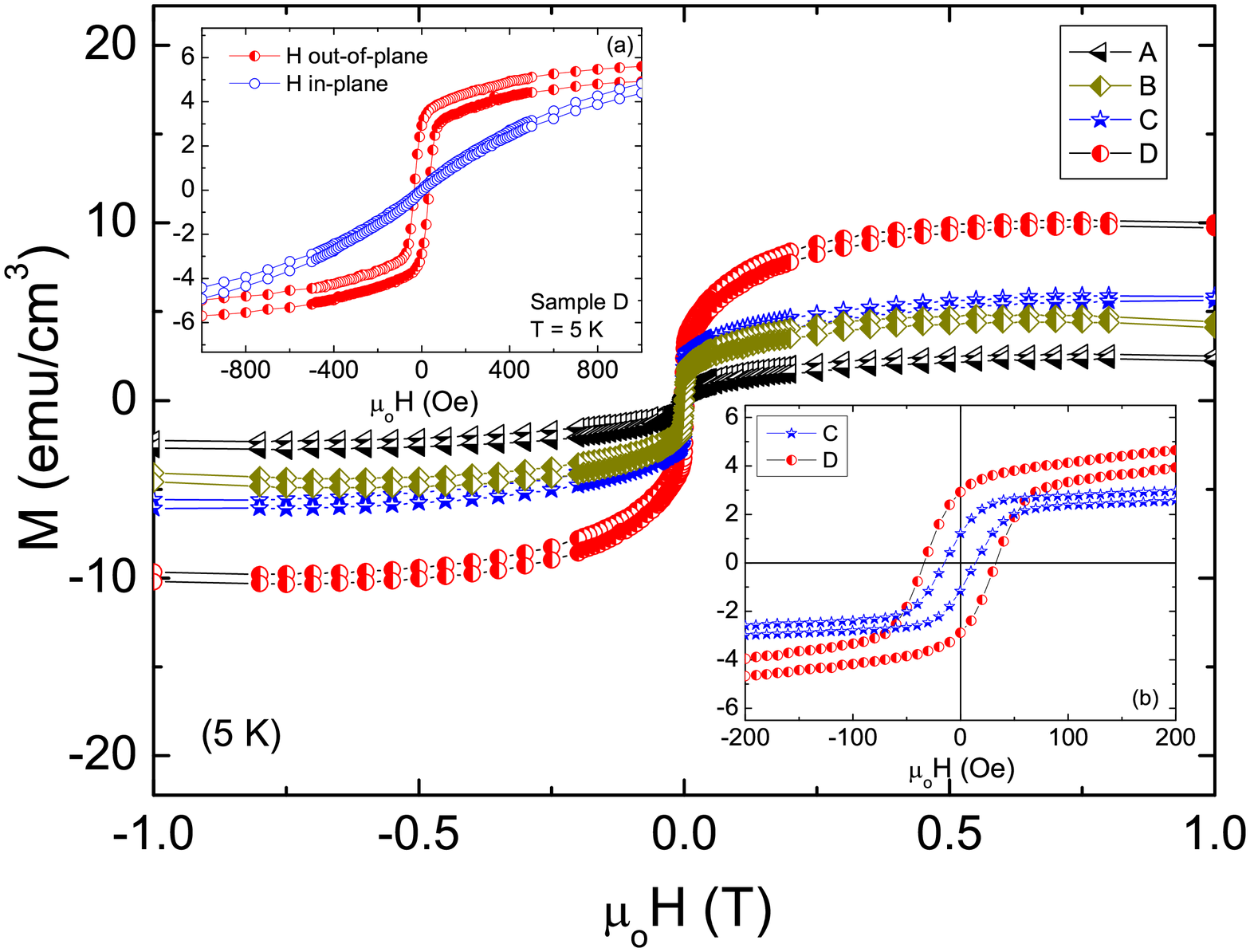}
\caption{\label{Fig.5} Magnetization as a function of applied field for several laser annealed InMnP samples (A, B, C and D). The saturation magnetization increases with the Mn concentration. Inset (a) represents a magnetic anisotropy in InMnP epilayers and it also confirms that the InMnP epilayer has an out-of-plane easy axis. Inset (b) shows the Mn concentration dependent remanence and coercive field in InMnP samples C and D.}
\end{center}
\end{figure}
Figure~\ref{Fig.5} shows the magnetization of InMnP samples A, B, C and D as a function of an applied magnetic field measured at 5 K. The magnetic signals of InMnP consists of a paramagnetic response at low temperature mainly due to the regions where the Mn ions do not order ferromagnetically, either due to structural disorder or due to low local Mn concentration. The magnetization of InMnP samples increases with Mn concentration and it is $\sim$ 10 emu/cm$^3$ or $\sim$1$\pm 0.2$ $\mu_B$/Mn for sample D. The value of saturation magnetization is largely underestimated due to several reasons. These include the sputtering effect in III-V semiconductors during ion implantation which reduces the effective implantation fluence \cite{Fritzsche39}, and the formation of a magnetically inert layer with a large Mn concentration due to surface segregation during the pulsed laser annealing \cite{Scarpulla073913}. Furthermore, a self-compensation process (interstitial Mn, antisites) as in GaMnAs \cite{edmonds2004mn,Lima012406} can further depress the ferromagnetism in InMnP. A magnetic anisotropy perpendicular to the sample plane e.g., the magnetic easy axis is out-of-plane, has been observed in our InMnP samples (see inset (a) to Fig.~\ref{Fig.5} for an example). It could be due to the following reason. The substitution of Indium by manganese ions results in a smaller lattice constant of InMnP compared to its bulk InP. Therefore, the InMnP layer is under a tensile strain. Due to the biaxial tensile strain, the valence band splits and the lowest valence band assumes a heavy-hole character \cite{dietl2001holePRB}. The hole spins are oriented along the growth direction when only the lowest valence band is occupied, since in this case it can lower their energy by coupling to the Mn spins and hence a perpendicular magnetic anisotropy is expected in this case \cite{Sawicki245325,thevenard2006magnetic,stone2010interplay,Rushforth073908}. The tensile strain in the InMnP layer is confirmed by XRD results and it increases with the Mn concentration. On the other hand, the magnetic anisotropy in ferromagnetic semiconductors also depends on temperature \cite{Sawicki245325} and on hole concentration \cite{sawicki2005plane,stone2008compensation,glunk2009magnetic,casiraghi2010tuning}. 

The coercive field in InMnP also varies with Mn concentration, it increases with the Mn concentration (see inset (b) to Fig.~\ref{Fig.5}). The increase in the coercive field in case of a high Mn concentration in InMnP may be attributed to defects acting as domain wall pinning sites.\newline
\begin{figure}[t]
\begin{center}
\includegraphics[width=0.46\textwidth]{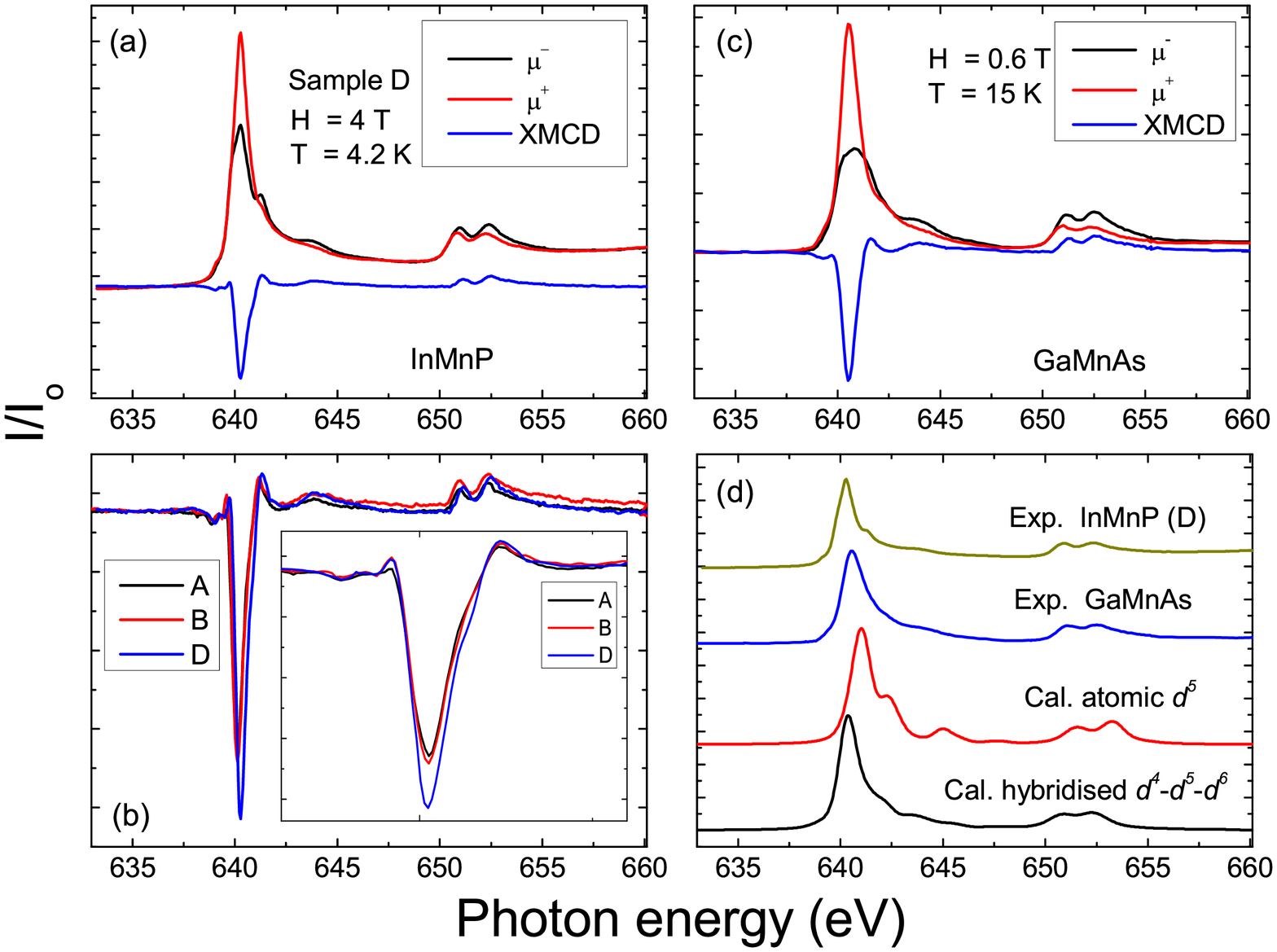}
\caption{\label{Fig.6} (a,c) A qualitative comparison of XAS and XMCD spectra of InMnP and GaMnAs measured at temperatures 4.2 K and 15 K, under fields of 4 T and 0.6 T, respectively. (b) XMCD difference spectra of InMnP samples containing different Mn concentrations. (d) Calculated XAS spectra for a localized Mn $d^5$ ground state and for a hybrid state (16$\%$ $d^4$, 58$\%$ $d^5$, 26$\%$ $d^6$) in GaAs along with experimental XAS spectra taken from InMnP and GaMnAs. Note that (c) and a part of (d) are reproduced under permission \cite{Edmonds4065}. All spectra in (D) are rescaled.}
\end{center}
\end{figure}
X-ray absorption spectroscopy (XAS) in combination with x-ray magnetic circular dichroism (XMCD) are widely used to probe the electronic and magnetic structures of matter. XMCD in particular provides information on a selected element of a specimen. We have performed XMCD measurements on our InMnP samples at the Mn L$_{2,3}$ edges. At these Mn edges the 2p core electrons are excited to the unoccupied 3d states and consequently information about the electronic structure of the polarized Mn band is obtained (see Ref. \cite{Schutz737} for more details). It is well known that an oxide surface layer is formed in Mn doped III-V semiconductors \cite{Edmonds4065,Stone012504}. Therefore, a HCl solution  is used to remove the oxide layer prior to XMCD measurements. For the XMCD measurements in this particular study, all InMnP samples are measured at the lowest possible temperature ($\sim$ 4.2 K) and under a magnetic field of 4 T.\newline
Figure~\ref{Fig.6}(a) represents the results of XAS and XMCD measurements of sample D taken at the Mn L$_{2,3}$ edges in a total electron yield (TEY) mode. The main strong features in the XAS spectra at the L$_{2,3}$ edges (at energies around 640 and 652 eV) are due to electric-dipole allowed transitions (i.e., $\Delta l = \pm 1$) from 2p to unfilled 3d states. For comparison, XAS and XMCD spectra of a GaMnAs film grown by molecular beam epitaxy are shown in Fig.~\ref{Fig.6}(c). At first glance, the qualitative features of XAS and XMCD spectra in both systems are quite similar which indicates that both systems have the same chemical environment of Mn species. Similar XAS spectral features are observed in other III-V semiconductors such as GaMnP and InMnAs \cite{Scarpulla073913,Stone012504,Zhou093007}. The calculated XAS spectra of GaMnAs at Mn L$_{2,3}$ edges for the atomic $d^5$ and hybridized Mn $d$ (Mn $d$ states are hybridized with As $p$ states in the valence band of GaMnAs) ground states are shown in Fig.~\ref{Fig.6}(d). Our measured XAS spectra of InMnP are similar to that of calculated XAS spectra of hybridized Mn ground states instead of localized atomic Mn ground states as shown in Fig.~\ref{Fig.6}(d). The similar XAS spectral features of Mn-doped III-V semiconductors hint to a similar bonding and exchange interaction in these materials even though they have different energy gaps and Mn-acceptor levels.\newline
The XMCD signal at the Mn L$_3$-edge of sample D defined as ($\mu^+$-$\mu^-$)/($\mu^+$+$\mu^-$) $\times$ 100 is $\sim$ $40\%$ at 4.2 K under a magnetic field of 4 T which decreases in samples containing a low Mn concentration (see table~\ref{table1}). This value of XMCD is comparable to that of other ferromagnetic III-V semiconductors \cite{Scarpulla073913,Zhou093007,Stone012504}. The large XMCD signal of InMnP also indicates that it has a high spin polarization at the Fermi energy (E$_f$) which depends on the Mn concentration, see Fig.~\ref{Fig.6}(b). The XMCD sum rules provide information on the degree of the spin and orbital moments in the system \cite{Thole1943,Obrien12672}. The spin moment calculated using XMCD data and the sum rules in sample D is $\sim$ 1$\pm 0.1$ $\mu_B$/Mn while the orbital moment is negligibly small.  

Both the Curie temperature and the XMCD at the Mn L-edge increase with Mn concentration. Indeed, the XMCD signal also shows a similar temperature dependent behavior as the magnetization measured by SQUID-VSM \cite{Khalid121301}. Both methods probe the same ferromagnetic phase. Therefore, we can prove that the ferromagnetism in InMnP is intrinsic and due to substituted Mn ions. 

\begin{figure}[t]
\begin{center}
\includegraphics[width=0.45\textwidth]{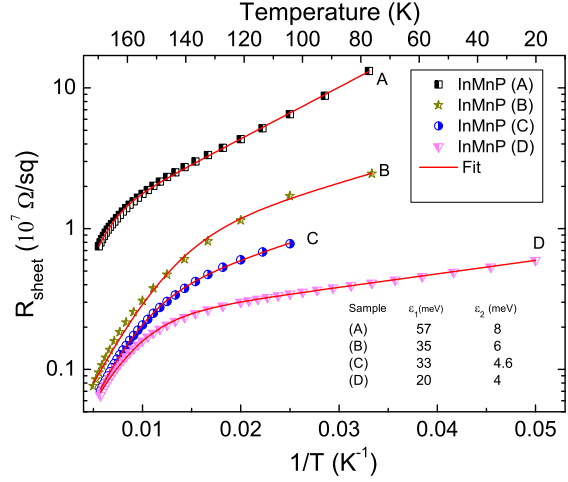}
\caption{\label{Fig.7} Sheet resistance of InMnP samples A, B, C and D as a function of inverse temperature. The red solid lines are fits of experimental data to Eq.~\ref{1}. Note that the top x-axis panel shows the standard linear scale for sample D.}
\end{center}
\end{figure}
\subsection{Transport properties}
The carrier mediated nature of ferromagnetism and the conduction mechanism in InMnP can also be studied by transport methods (magnetoresistance, anomalous Hall effect and resistivity). The four samples A, B, C, D were used for magneto-transport measurements. We carried out temperature dependent resistivity and magnetoresitance measurements using a van der Pauw geometery under a magnetic field perpendicular to the sample plane. Figure~\ref{Fig.7} represents the sheet resistance as a function of inverse temperature for samples A, B, C and D measured under a zero magnetic field. All samples exhibit semiconductor like sheet resistance up to the Mn concentration as high as 5 at.$\%$ and to the lowest measurable temperature. The temperature dependent sheet resistance of InMnP is quite similar to that of GaMnP \cite{Scarpulla207204}, but different from that of GaMnAs with comparable Mn concentration \cite{MatsukuraR2037}. It is worth noting that two different slopes, at low and high temperatures, in the temperature dependent sheet resistance of all four samples hint to different conduction mechanisms at low and high temperatures. Therefore, to describe thermally activated conduction processes at low and high temperatures, we have used a model \cite{Scarpulla207204} which is given in Eq.~\ref{1},
\begin{equation}
\label{1}
\rho(T)^{-1}=\left\{\rho_1\exp(E_1/\kappa_BT)\right\}^{-1}+\left\{\rho_2\exp(E_2/\kappa_BT)\right\}^{-1}
\end{equation}
where pre-exponential constants $\rho_1$, $\rho_2$ and activation energies $E_1$, $E_2$ are the free parameters. This model fits quite nicely to the measured temperature dependent sheet resistance of all InMnP samples. The fitting results indicate two thermally activated contributions to the resistivity of the samples. A high temperature activation region, where mainly electron transition between the valence band and the acceptor states occurs. A low temperature activation region, where the hopping conduction dominates in InMnP which has also been reported theoretically in other III-V:Mn semiconductors \cite{Kaminski235210}. The high and low activation energies $E_1$ and $E_2$ for our samples A, B, C, D are 57, 35, 33, 20 and 8, 6, 4.6, 4 meV, respectively. For InMnP the activation energy is smaller when compared with GaMnP with similar Mn concentration \cite{Scarpulla207204}. This was expected due to the following reasons: first, at high temperatures the low resistivity in InMnP is mainly due to the transition between the valence band and the acceptor states. Second, the Mn isolated acceptor state in InMnP lies at a low energy (220 meV) compared to GaMnP (400 meV), therefore, less energy is needed to excite an electron from the valence band to the acceptor states. A decrease in the activation energy of InMnP with the Mn concentration was expected due to the band broadening at high Mn doping. A close examination of the resistivity data of InMnP samples in Fig.~\ref{Fig.7} also indicates a continuous change in the slope near the Curie temperatures, which can be due to the nearest-neighbor hopping transport \cite{Kaminski235210} at low temperatures. The activation energy for a low Mn-doped sample A increases due to the narrowing of the impurity band and the shift of the Fermi energy in the Mn-induced band accompanied by a reduction in hole concentration. The high insulating character even at 5 at.$\%$ Mn doping and the hopping transport in InMnP support a detached impurity band in InMnP.\\     
\newline
\begin{figure}[t]
\begin{center}
\includegraphics[width=0.4\textwidth]{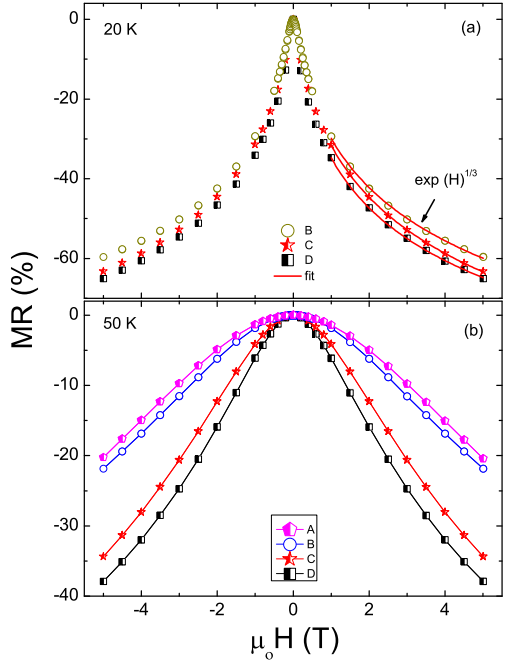}
\caption{\label{Fig.8} Magnetoresistance of InMnP samples measured at temperatures (a) 20 K and (b) 50 K. The red solid lines in (a) show the fits of the data to a model given in Eq.~\ref{2}.}
\end{center}
\end{figure}
We have performed magnetoresistance (MR) measurements on samples A, B, C and D in order to investigate a magnetic field response of the resistivity. Particulary, the low temperature magnetoresistance data also provides information on the transport mechansim in the system. Figure~\ref{Fig.8} (a) shows the magnetoresistance of three InMnP samples B, C and D measured at 20 K under a magnetic field of 5 T, while Fig.~\ref{Fig.8} (b) shows the MR of InMnP samples above the Curie temperature. The curvature of a MR curve above and below T$_c$ is different which indicates two different kinds of transport mechanisms in this system. However, all samples exhibit negative magnetoresistance which is typical for III-V dilute magnetic semiconductors. The relative magnetoresistance defined as $MR(\%)=\left[\left\{\rho(H)-\rho(0)\right\}/\rho(0)\right]\times100$ for sample B is about 59$\%$ at 5 T which increases with the Mn concentration and reaches a value of 65$\%$ at a Mn concentration of 5 at.$\%$ (see table~\ref{table1}). The negative magnetoresistance in InMnP can be explained as follows \cite{Esch13103}: when a magnetic field is applied, it results in an antiferromagnetic coupling between the Mn ion and the hole (a ferromagnetic one between two Mn ions). Consequently, the wavefunctions of Mn-hole complexes expand and the increased overlapping of these wavefunctions results in a negative magnetoresistance of InMnP. We have used a model in high-field limits i.e. $\lambda$$\ll$$a$, where $\lambda$ and $a$ are the magnetic length and localization radius, respectively, which has been used for insulating GaMnAs previously \cite{Esch13103}. The model is given in Eq.~\ref{2}. 
\begin{equation}
\label{2}
\rho(H)=\rho_o\exp\left[\frac{C}{(\lambda^2T)^{1/3}}\right]
\end{equation}
\begin{equation}
\label{2a}
\lambda=\left(\frac{c\hbar}{e\mu_0 H}\right)^{1/2}
\end{equation}
where $\lambda$ is the magnetic length. This model is used to fit the magnetoresistance data at low temperature for three InMnP samples and the results are shown in Fig.~\ref{Fig.8}(a). The fit results show that the model is applicable to explain the magnetoresistance of InMnP in the high-field range of 1-5 T. The fitting parameter $C$ has a negative sign due to the expansion of the wave functions in a magnetic field which increases with the Mn concentration from 12910 for sample B to 20465 for sample D (see table~\ref{table1}). The dependence of the fitting parameter $C$ on the Mn concentration indicates that under a magnetic field the overlapping of the wave functions of Mn-h complexes increases which results in a large negative magnetoresistance of sample D. At low temperature the magnetization reaches its saturation, the decrease in resistivity under a magnetic field is very weak. As at low temperature there remains certain conductivity in InMnP samples which could be related to the hopping conduction in the impurity band. The high insulating characteristics of InMnP samples (low mobility) also indicate that the Fermi level resides in the localized impurity band. In this scenario, the hopping conduction is dominating in the impurity band because it needs very low energy (few meV). Therefore, we can conclude that hopping within the impurity band is the main conduction mechanism in InMnP at low temperatures which is also an indication of a separated Mn-impurity induced band within the bandgap of InP.
\section{Conclusion}
We have prepared a dilute ferromagnetic semiconductor InMnP by Mn ion implantation and pulsed laser annealing. Transmission electron microscopy with a combination of electron diffraction is used to study the microstructural changes produced by Mn ions in InP. From the micrographs it is seen that InP samples become almost amorphous after Mn implantation but recrystallize after pulsed laser annealing. The thickness of the InMnP epilayers is found to be 95$\pm$5 nm as estimated by SRIM and confirmed by HRTEM and XRD results. The Curie temperature of InMnP samples depends on the Mn concentration and reaches to 40$\pm$2 K when the Mn concentration is as high as 5 at.$\%$. The shape of the M-T curves do not follow the usual mean-field theory instead they reflect a strong compensation contribution of Mn ions in InMnP epilayers. The saturation magnetization depends on the Mn concentration and reaches to 10 emu/cm$^3$ for sample D.\newline The large XMCD signal in InMnP samples reflects a strong spin polarization at the Fermi level in this system. A comparison of XAS or XMCD results obtained from InMnP and GaMnAs compounds indicates that Mn ion has a hybridized ground state (Mn$^{2+}$) in InMnP as in GaMnAs.\newline 
However, transport results suggest that inspite of a similar chemical environment in III-V semiconductors (InMnP, GaMnP, GaMnP, InMnAs), the degree of Mn ions incorporation in these compounds is different. These compounds, therefore, have different electrical and magnetic properties. The transport mechanism in InMnP is investigated by varying the Mn concentration. We did not observe an insulator-metal transition in InMnP up to a Mn concentration of 5 at.$\%$ instead all InMnP samples show insulating characteristics. Magneotresistance results obtained at low temperatures support the hopping conduction mechanism in InMnP. We find that the Mn impurity band remains detached from the valence band in InMnP up to 5 at.$\%$ Mn doping. Our findings indicate that the local environement of Mn ions in InMnP is similar to those of GaMnAs, GaMnP and InMnAs. It also seems that an unmerged Mn impurity band is formed in the bandgap of InP. This work might be helpful in understanding the family of III-V:Mn semiconductors, i.e., the different Mn binding energy in different III-V compounds should be considered.\newline 
\\
The authors thank Stefan Facsko for AES measurements. The work is financially supported by the Helmholtz-Gemeinschaft Deutscher Forschungszentren (VH-NG-713).  

\begin{thebibliography}{52}
\expandafter\ifx\csname natexlab\endcsname\relax\def\natexlab#1{#1}\fi
\expandafter\ifx\csname bibnamefont\endcsname\relax
  \def\bibnamefont#1{#1}\fi
\expandafter\ifx\csname bibfnamefont\endcsname\relax
  \def\bibfnamefont#1{#1}\fi
\expandafter\ifx\csname citenamefont\endcsname\relax
  \def\citenamefont#1{#1}\fi
\expandafter\ifx\csname url\endcsname\relax
  \def\url#1{\texttt{#1}}\fi
\expandafter\ifx\csname urlprefix\endcsname\relax\def\urlprefix{URL }\fi
\providecommand{\bibinfo}[2]{#2}
\providecommand{\eprint}[2][]{\url{#2}}

\bibitem[{\citenamefont{Ohno et~al.}(1999)\citenamefont{Ohno, Young, Beschoten,
  Matsukura, Ohno, and Awschalom}}]{Ohno790}
\bibinfo{author}{\bibfnamefont{Y.}~\bibnamefont{Ohno}},
  \bibinfo{author}{\bibfnamefont{D.}~\bibnamefont{Young}},
  \bibinfo{author}{\bibfnamefont{B.}~\bibnamefont{Beschoten}},
  \bibinfo{author}{\bibfnamefont{F.}~\bibnamefont{Matsukura}},
  \bibinfo{author}{\bibfnamefont{H.}~\bibnamefont{Ohno}}, \bibnamefont{and}
  \bibinfo{author}{\bibfnamefont{D.}~\bibnamefont{Awschalom}},
  \bibinfo{journal}{Nature} \textbf{\bibinfo{volume}{402}},
  \bibinfo{pages}{790} (\bibinfo{year}{1999}).

\bibitem[{\citenamefont{Ohno et~al.}(2000)\citenamefont{Ohno, Chiba, Matsukura,
  Omiya, Abe, Dietl, Ohno, and Ohtani}}]{Ohno944}
\bibinfo{author}{\bibfnamefont{H.}~\bibnamefont{Ohno}},
  \bibinfo{author}{\bibfnamefont{D.}~\bibnamefont{Chiba}},
  \bibinfo{author}{\bibfnamefont{F.}~\bibnamefont{Matsukura}},
  \bibinfo{author}{\bibfnamefont{T.}~\bibnamefont{Omiya}},
  \bibinfo{author}{\bibfnamefont{E.}~\bibnamefont{Abe}},
  \bibinfo{author}{\bibfnamefont{T.}~\bibnamefont{Dietl}},
  \bibinfo{author}{\bibfnamefont{Y.}~\bibnamefont{Ohno}}, \bibnamefont{and}
  \bibinfo{author}{\bibfnamefont{K.}~\bibnamefont{Ohtani}},
  \bibinfo{journal}{Nature} \textbf{\bibinfo{volume}{408}},
  \bibinfo{pages}{944} (\bibinfo{year}{2000}).

\bibitem[{\citenamefont{Chiba et~al.}(2008)\citenamefont{Chiba, Sawicki,
  Nishitani, Nakatani, Matsukura, and Ohno}}]{Chiba515}
\bibinfo{author}{\bibfnamefont{D.}~\bibnamefont{Chiba}},
  \bibinfo{author}{\bibfnamefont{M.}~\bibnamefont{Sawicki}},
  \bibinfo{author}{\bibfnamefont{Y.}~\bibnamefont{Nishitani}},
  \bibinfo{author}{\bibfnamefont{Y.}~\bibnamefont{Nakatani}},
  \bibinfo{author}{\bibfnamefont{F.}~\bibnamefont{Matsukura}},
  \bibnamefont{and} \bibinfo{author}{\bibfnamefont{H.}~\bibnamefont{Ohno}},
  \bibinfo{journal}{Nature} \textbf{\bibinfo{volume}{455}},
  \bibinfo{pages}{515} (\bibinfo{year}{2008}).

\bibitem[{\citenamefont{Scarpulla et~al.}(2005)\citenamefont{Scarpulla,
  Cardozo, Farshchi, Oo, McCluskey, Yu, and Dubon}}]{Scarpulla207204}
\bibinfo{author}{\bibfnamefont{M.}~\bibnamefont{Scarpulla}},
  \bibinfo{author}{\bibfnamefont{B.}~\bibnamefont{Cardozo}},
  \bibinfo{author}{\bibfnamefont{R.}~\bibnamefont{Farshchi}},
  \bibinfo{author}{\bibfnamefont{W.~H.} \bibnamefont{Oo}},
  \bibinfo{author}{\bibfnamefont{M.}~\bibnamefont{McCluskey}},
  \bibinfo{author}{\bibfnamefont{K.}~\bibnamefont{Yu}}, \bibnamefont{and}
  \bibinfo{author}{\bibfnamefont{O.}~\bibnamefont{Dubon}},
  \bibinfo{journal}{Phys. Rev. Lett.} \textbf{\bibinfo{volume}{95}},
  \bibinfo{pages}{207204} (\bibinfo{year}{2005}).

\bibitem[{\citenamefont{Wolf et~al.}(2001)\citenamefont{Wolf, Awschalom,
  Buhrman, Daughton, von Molnár, Roukes, Chtchelkanova, and
  Treger}}]{Wolf1488}
\bibinfo{author}{\bibfnamefont{S.}~\bibnamefont{Wolf}},
  \bibinfo{author}{\bibfnamefont{D.}~\bibnamefont{Awschalom}},
  \bibinfo{author}{\bibfnamefont{R.}~\bibnamefont{Buhrman}},
  \bibinfo{author}{\bibfnamefont{J.}~\bibnamefont{Daughton}},
  \bibinfo{author}{\bibfnamefont{S.}~\bibnamefont{von Molnár}},
  \bibinfo{author}{\bibfnamefont{M.}~\bibnamefont{Roukes}},
  \bibinfo{author}{\bibfnamefont{A.}~\bibnamefont{Chtchelkanova}},
  \bibnamefont{and} \bibinfo{author}{\bibfnamefont{D.}~\bibnamefont{Treger}},
  \bibinfo{journal}{Science} \textbf{\bibinfo{volume}{294}},
  \bibinfo{pages}{1488} (\bibinfo{year}{2001}).

\bibitem[{\citenamefont{Dietl et~al.}(2000)\citenamefont{Dietl, Ohno,
  Matsukura, Cibert, and Ferrand}}]{Dietl1019}
\bibinfo{author}{\bibfnamefont{T.}~\bibnamefont{Dietl}},
  \bibinfo{author}{\bibfnamefont{H.}~\bibnamefont{Ohno}},
  \bibinfo{author}{\bibfnamefont{F.}~\bibnamefont{Matsukura}},
  \bibinfo{author}{\bibfnamefont{J.}~\bibnamefont{Cibert}}, \bibnamefont{and}
  \bibinfo{author}{\bibfnamefont{D.}~\bibnamefont{Ferrand}},
  \bibinfo{journal}{Science} \textbf{\bibinfo{volume}{287}},
  \bibinfo{pages}{1019} (\bibinfo{year}{2000}).

\bibitem[{\citenamefont{Ma\v{s}ek et~al.}(2010)\citenamefont{Ma\v{s}ek, M\'aca,
  Kudrnovsk\'y, Makarovsky, Eaves, Campion, Edmonds, Rushforth, Foxon,
  Gallagher et~al.}}]{Masek227202}
\bibinfo{author}{\bibfnamefont{J.}~\bibnamefont{Ma\v{s}ek}},
  \bibinfo{author}{\bibfnamefont{F.}~\bibnamefont{M\'aca}},
  \bibinfo{author}{\bibfnamefont{J.}~\bibnamefont{Kudrnovsk\'y}},
  \bibinfo{author}{\bibfnamefont{O.}~\bibnamefont{Makarovsky}},
  \bibinfo{author}{\bibfnamefont{L.}~\bibnamefont{Eaves}},
  \bibinfo{author}{\bibfnamefont{R.}~\bibnamefont{Campion}},
  \bibinfo{author}{\bibfnamefont{K.}~\bibnamefont{Edmonds}},
  \bibinfo{author}{\bibfnamefont{A.}~\bibnamefont{Rushforth}},
  \bibinfo{author}{\bibfnamefont{C.}~\bibnamefont{Foxon}},
  \bibinfo{author}{\bibfnamefont{B.}~\bibnamefont{Gallagher}},
  \bibnamefont{et~al.}, \bibinfo{journal}{Phys. Rev. Lett.}
  \textbf{\bibinfo{volume}{105}}, \bibinfo{pages}{227202}
  (\bibinfo{year}{2010}).

\bibitem[{\citenamefont{Burch et~al.}(2006)\citenamefont{Burch, Shrekenhamer,
  Singley, Stephens, Sheu, Kawakami, Schiffer, Samarth, Awschalom, and
  Basov}}]{Burch087208}
\bibinfo{author}{\bibfnamefont{K.}~\bibnamefont{Burch}},
  \bibinfo{author}{\bibfnamefont{D.}~\bibnamefont{Shrekenhamer}},
  \bibinfo{author}{\bibfnamefont{E.}~\bibnamefont{Singley}},
  \bibinfo{author}{\bibfnamefont{J.}~\bibnamefont{Stephens}},
  \bibinfo{author}{\bibfnamefont{B.}~\bibnamefont{Sheu}},
  \bibinfo{author}{\bibfnamefont{R.}~\bibnamefont{Kawakami}},
  \bibinfo{author}{\bibfnamefont{P.}~\bibnamefont{Schiffer}},
  \bibinfo{author}{\bibfnamefont{N.}~\bibnamefont{Samarth}},
  \bibinfo{author}{\bibfnamefont{D.}~\bibnamefont{Awschalom}},
  \bibnamefont{and} \bibinfo{author}{\bibfnamefont{D.}~\bibnamefont{Basov}},
  \bibinfo{journal}{Phys. Rev. Lett.} \textbf{\bibinfo{volume}{97}},
  \bibinfo{pages}{087208} (\bibinfo{year}{2006}).

\bibitem[{\citenamefont{Hirakawa et~al.}(2002)\citenamefont{Hirakawa,
  Katsumoto, Hayashi, Hashimoto, and Iye}}]{Hirakawa193312}
\bibinfo{author}{\bibfnamefont{K.}~\bibnamefont{Hirakawa}},
  \bibinfo{author}{\bibfnamefont{S.}~\bibnamefont{Katsumoto}},
  \bibinfo{author}{\bibfnamefont{T.}~\bibnamefont{Hayashi}},
  \bibinfo{author}{\bibfnamefont{Y.}~\bibnamefont{Hashimoto}},
  \bibnamefont{and} \bibinfo{author}{\bibfnamefont{Y.}~\bibnamefont{Iye}},
  \bibinfo{journal}{Phys. Rev. B} \textbf{\bibinfo{volume}{65}},
  \bibinfo{pages}{193312} (\bibinfo{year}{2002}).

\bibitem[{\citenamefont{Ando et~al.}(2008)\citenamefont{Ando, Saito, Agarwal,
  Debnath, and Zayets}}]{Ando067204}
\bibinfo{author}{\bibfnamefont{K.}~\bibnamefont{Ando}},
  \bibinfo{author}{\bibfnamefont{H.}~\bibnamefont{Saito}},
  \bibinfo{author}{\bibfnamefont{K.}~\bibnamefont{Agarwal}},
  \bibinfo{author}{\bibfnamefont{M.}~\bibnamefont{Debnath}}, \bibnamefont{and}
  \bibinfo{author}{\bibfnamefont{V.}~\bibnamefont{Zayets}},
  \bibinfo{journal}{Phys. Rev. Lett.} \textbf{\bibinfo{volume}{100}},
  \bibinfo{pages}{067204} (\bibinfo{year}{2008}).

\bibitem[{\citenamefont{Rokhinson et~al.}(2007)\citenamefont{Rokhinson,
  Lyanda-Geller, Ge, Shen, Liu, Dobrowolska, and Furdyna}}]{Rokhinson161201}
\bibinfo{author}{\bibfnamefont{L.}~\bibnamefont{Rokhinson}},
  \bibinfo{author}{\bibfnamefont{Y.}~\bibnamefont{Lyanda-Geller}},
  \bibinfo{author}{\bibfnamefont{Z.}~\bibnamefont{Ge}},
  \bibinfo{author}{\bibfnamefont{S.}~\bibnamefont{Shen}},
  \bibinfo{author}{\bibfnamefont{X.}~\bibnamefont{Liu}},
  \bibinfo{author}{\bibfnamefont{M.}~\bibnamefont{Dobrowolska}},
  \bibnamefont{and} \bibinfo{author}{\bibfnamefont{J.}~\bibnamefont{Furdyna}},
  \bibinfo{journal}{Phys. Rev. B} \textbf{\bibinfo{volume}{76}},
  \bibinfo{pages}{161201(R)} (\bibinfo{year}{2007}).

\bibitem[{\citenamefont{Ohya et~al.}(2011)\citenamefont{Ohya, Takata, and
  Tanaka}}]{Ohya342}
\bibinfo{author}{\bibfnamefont{S.}~\bibnamefont{Ohya}},
  \bibinfo{author}{\bibfnamefont{K.}~\bibnamefont{Takata}}, \bibnamefont{and}
  \bibinfo{author}{\bibfnamefont{M.}~\bibnamefont{Tanaka}},
  \bibinfo{journal}{Nature Phys.} \textbf{\bibinfo{volume}{7}},
  \bibinfo{pages}{342} (\bibinfo{year}{2011}).

\bibitem[{\citenamefont{Dobrowolska et~al.}(2012)\citenamefont{Dobrowolska,
  Tivakornsasithorn, Liu, Furdyna, Berciu, M.Yu, and
  Walukiewicz}}]{Dobrowolska444}
\bibinfo{author}{\bibfnamefont{M.}~\bibnamefont{Dobrowolska}},
  \bibinfo{author}{\bibfnamefont{K.}~\bibnamefont{Tivakornsasithorn}},
  \bibinfo{author}{\bibfnamefont{X.}~\bibnamefont{Liu}},
  \bibinfo{author}{\bibfnamefont{J.}~\bibnamefont{Furdyna}},
  \bibinfo{author}{\bibfnamefont{M.}~\bibnamefont{Berciu}},
  \bibinfo{author}{\bibfnamefont{K.}~\bibnamefont{M.Yu}}, \bibnamefont{and}
  \bibinfo{author}{\bibfnamefont{W.}~\bibnamefont{Walukiewicz}},
  \bibinfo{journal}{Nature Materials} \textbf{\bibinfo{volume}{11}},
  \bibinfo{pages}{444} (\bibinfo{year}{2012}).

\bibitem[{\citenamefont{Clerjaud}(1985)}]{Clerjaud3615}
\bibinfo{author}{\bibfnamefont{B.}~\bibnamefont{Clerjaud}},
  \bibinfo{journal}{J.Phys.C:Solid State Phys.} \textbf{\bibinfo{volume}{18}},
  \bibinfo{pages}{3615} (\bibinfo{year}{1985}).

\bibitem[{\citenamefont{Ye et~al.}(2014)\citenamefont{Ye, Wang, Khalid, Gao,
  Prucnal, Gordan, Salvan, Zahn, Skorupa, Helm et~al.}}]{yuan2401304}
\bibinfo{author}{\bibfnamefont{Y.}~\bibnamefont{Ye}},
  \bibinfo{author}{\bibfnamefont{Y.}~\bibnamefont{Wang}},
  \bibinfo{author}{\bibfnamefont{M.}~\bibnamefont{Khalid}},
  \bibinfo{author}{\bibfnamefont{K.}~\bibnamefont{Gao}},
  \bibinfo{author}{\bibfnamefont{S.}~\bibnamefont{Prucnal}},
  \bibinfo{author}{\bibfnamefont{O.}~\bibnamefont{Gordan}},
  \bibinfo{author}{\bibfnamefont{G.}~\bibnamefont{Salvan}},
  \bibinfo{author}{\bibfnamefont{D.}~\bibnamefont{Zahn}},
  \bibinfo{author}{\bibfnamefont{W.}~\bibnamefont{Skorupa}},
  \bibinfo{author}{\bibfnamefont{M.}~\bibnamefont{Helm}}, \bibnamefont{et~al.},
  \bibinfo{journal}{IEEE TRANSACTIONS ON MAGNETICS}
  \textbf{\bibinfo{volume}{50}}, \bibinfo{pages}{2401304}
  (\bibinfo{year}{2014}).

\bibitem[{\citenamefont{Khalid et~al.}(2014)\citenamefont{Khalid, Weschke,
  Skorupa, Helm, and Zhou}}]{Khalid121301}
\bibinfo{author}{\bibfnamefont{M.}~\bibnamefont{Khalid}},
  \bibinfo{author}{\bibfnamefont{E.}~\bibnamefont{Weschke}},
  \bibinfo{author}{\bibfnamefont{W.}~\bibnamefont{Skorupa}},
  \bibinfo{author}{\bibfnamefont{M.}~\bibnamefont{Helm}}, \bibnamefont{and}
  \bibinfo{author}{\bibfnamefont{S.}~\bibnamefont{Zhou}},
  \bibinfo{journal}{Phys. Rev. B} \textbf{\bibinfo{volume}{89}},
  \bibinfo{pages}{121301(R)} (\bibinfo{year}{2014}).

\bibitem[{\citenamefont{Ziegler}(2003)}]{Ziegler1027}
\bibinfo{author}{\bibfnamefont{J.}~\bibnamefont{Ziegler}},
  \bibinfo{journal}{Nucl. Instrum. Methods}
  \textbf{\bibinfo{volume}{B219/220}}, \bibinfo{pages}{1027}
  (\bibinfo{year}{2003}).

\bibitem[{\citenamefont{Scarpulla}(2006)}]{Dissertation145}
\bibinfo{author}{\bibfnamefont{M.}~\bibnamefont{Scarpulla}},
  \bibinfo{journal}{University of California, Berkeley} p. \bibinfo{pages}{145}
  (\bibinfo{year}{2006}).

\bibitem[{\citenamefont{Wu et~al.}(2004)\citenamefont{Wu, Zhou, Yao, Zhao,
  Vantomme, Daele, Piscopiello, Tendeloo, Tong, Yang et~al.}}]{Wu920}
\bibinfo{author}{\bibfnamefont{M.}~\bibnamefont{Wu}},
  \bibinfo{author}{\bibfnamefont{S.}~\bibnamefont{Zhou}},
  \bibinfo{author}{\bibfnamefont{S.}~\bibnamefont{Yao}},
  \bibinfo{author}{\bibfnamefont{Q.}~\bibnamefont{Zhao}},
  \bibinfo{author}{\bibfnamefont{A.}~\bibnamefont{Vantomme}},
  \bibinfo{author}{\bibfnamefont{B.~V.} \bibnamefont{Daele}},
  \bibinfo{author}{\bibfnamefont{E.}~\bibnamefont{Piscopiello}},
  \bibinfo{author}{\bibfnamefont{G.~V.} \bibnamefont{Tendeloo}},
  \bibinfo{author}{\bibfnamefont{Y.}~\bibnamefont{Tong}},
  \bibinfo{author}{\bibfnamefont{Z.}~\bibnamefont{Yang}}, \bibnamefont{et~al.},
  \bibinfo{journal}{J. Vac. Sci. Technol. B} \textbf{\bibinfo{volume}{22}},
  \bibinfo{pages}{920} (\bibinfo{year}{2004}).

\bibitem[{\citenamefont{Zhao et~al.}(2005)\citenamefont{Zhao, Staddon, Wang,
  Edmonds, Campion, Gallagher, and Foxon}}]{zhao2005intrinsic}
\bibinfo{author}{\bibfnamefont{L.}~\bibnamefont{Zhao}},
  \bibinfo{author}{\bibfnamefont{C.}~\bibnamefont{Staddon}},
  \bibinfo{author}{\bibfnamefont{K.}~\bibnamefont{Wang}},
  \bibinfo{author}{\bibfnamefont{K.}~\bibnamefont{Edmonds}},
  \bibinfo{author}{\bibfnamefont{R.}~\bibnamefont{Campion}},
  \bibinfo{author}{\bibfnamefont{B.}~\bibnamefont{Gallagher}},
  \bibnamefont{and} \bibinfo{author}{\bibfnamefont{C.}~\bibnamefont{Foxon}},
  \bibinfo{journal}{Appl. Phys. Lett.} \textbf{\bibinfo{volume}{86}},
  \bibinfo{pages}{071902} (\bibinfo{year}{2005}).

\bibitem[{\citenamefont{Yu and Cardona}(2010)}]{Yu375}
\bibinfo{author}{\bibfnamefont{P.~Y.} \bibnamefont{Yu}} \bibnamefont{and}
  \bibinfo{author}{\bibfnamefont{M.}~\bibnamefont{Cardona}},
  \emph{\bibinfo{title}{Fundamentals of Semiconductors}}
  (\bibinfo{publisher}{Springer Heidelberg}, \bibinfo{year}{2010}), p.
  \bibinfo{pages}{375}.

\bibitem[{\citenamefont{Trallero-Giner
  et~al.}(1989)\citenamefont{Trallero-Giner, Cantarero, and
  Cardona}}]{Trallero4030}
\bibinfo{author}{\bibfnamefont{C.}~\bibnamefont{Trallero-Giner}},
  \bibinfo{author}{\bibfnamefont{A.}~\bibnamefont{Cantarero}},
  \bibnamefont{and} \bibinfo{author}{\bibfnamefont{M.}~\bibnamefont{Cardona}},
  \bibinfo{journal}{Phys. Rev. B} \textbf{\bibinfo{volume}{40}},
  \bibinfo{pages}{4030} (\bibinfo{year}{1989}).

\bibitem[{\citenamefont{Seong et~al.}(2002)\citenamefont{Seong, Chun, Cheong,
  Samarth, and Mascarenhas}}]{Seong033202}
\bibinfo{author}{\bibfnamefont{M.}~\bibnamefont{Seong}},
  \bibinfo{author}{\bibfnamefont{S.}~\bibnamefont{Chun}},
  \bibinfo{author}{\bibfnamefont{H.}~\bibnamefont{Cheong}},
  \bibinfo{author}{\bibfnamefont{N.}~\bibnamefont{Samarth}}, \bibnamefont{and}
  \bibinfo{author}{\bibfnamefont{A.}~\bibnamefont{Mascarenhas}},
  \bibinfo{journal}{Phys. Rev. B} \textbf{\bibinfo{volume}{66}},
  \bibinfo{pages}{033202} (\bibinfo{year}{2002}).

\bibitem[{\citenamefont{Limmer et~al.}(2002)\citenamefont{Limmer, Glunk,
  Mascheck, Koeder, Klarer, Schoch, Thonke, Sauer, and Waag}}]{Limmer205209}
\bibinfo{author}{\bibfnamefont{W.}~\bibnamefont{Limmer}},
  \bibinfo{author}{\bibfnamefont{M.}~\bibnamefont{Glunk}},
  \bibinfo{author}{\bibfnamefont{S.}~\bibnamefont{Mascheck}},
  \bibinfo{author}{\bibfnamefont{A.}~\bibnamefont{Koeder}},
  \bibinfo{author}{\bibfnamefont{D.}~\bibnamefont{Klarer}},
  \bibinfo{author}{\bibfnamefont{W.}~\bibnamefont{Schoch}},
  \bibinfo{author}{\bibfnamefont{K.}~\bibnamefont{Thonke}},
  \bibinfo{author}{\bibfnamefont{R.}~\bibnamefont{Sauer}}, \bibnamefont{and}
  \bibinfo{author}{\bibfnamefont{A.}~\bibnamefont{Waag}},
  \bibinfo{journal}{Phys. Rev. B} \textbf{\bibinfo{volume}{66}},
  \bibinfo{pages}{205209} (\bibinfo{year}{2002}).

\bibitem[{\citenamefont{Maslar et~al.}(1993)\citenamefont{Maslar, Bohn,
  Ballegeer, Andideh, Adesida, Caneau, and Bhat}}]{Maslar2983}
\bibinfo{author}{\bibfnamefont{J.}~\bibnamefont{Maslar}},
  \bibinfo{author}{\bibfnamefont{P.}~\bibnamefont{Bohn}},
  \bibinfo{author}{\bibfnamefont{D.}~\bibnamefont{Ballegeer}},
  \bibinfo{author}{\bibfnamefont{E.}~\bibnamefont{Andideh}},
  \bibinfo{author}{\bibfnamefont{I.}~\bibnamefont{Adesida}},
  \bibinfo{author}{\bibfnamefont{C.}~\bibnamefont{Caneau}}, \bibnamefont{and}
  \bibinfo{author}{\bibfnamefont{R.}~\bibnamefont{Bhat}}, \bibinfo{journal}{J.
  Appl. Phys.} \textbf{\bibinfo{volume}{73}}, \bibinfo{pages}{2983}
  (\bibinfo{year}{1993}).

\bibitem[{\citenamefont{Yuasa and Ishii}(1987)}]{Yuasa3962}
\bibinfo{author}{\bibfnamefont{T.}~\bibnamefont{Yuasa}} \bibnamefont{and}
  \bibinfo{author}{\bibfnamefont{M.}~\bibnamefont{Ishii}},
  \bibinfo{journal}{Phys. Rev. B} \textbf{\bibinfo{volume}{35}},
  \bibinfo{pages}{3962} (\bibinfo{year}{1987}).

\bibitem[{\citenamefont{Ohno et~al.}(1992)\citenamefont{Ohno, Munekata, Penney,
  von Moln\'ar, and Chang}}]{Ohno2664}
\bibinfo{author}{\bibfnamefont{H.}~\bibnamefont{Ohno}},
  \bibinfo{author}{\bibfnamefont{H.}~\bibnamefont{Munekata}},
  \bibinfo{author}{\bibfnamefont{T.}~\bibnamefont{Penney}},
  \bibinfo{author}{\bibfnamefont{S.}~\bibnamefont{von Moln\'ar}},
  \bibnamefont{and} \bibinfo{author}{\bibfnamefont{L.~L.} \bibnamefont{Chang}},
  \bibinfo{journal}{Phys. Rev. Lett.} \textbf{\bibinfo{volume}{68}},
  \bibinfo{pages}{2664} (\bibinfo{year}{1992}).

\bibitem[{\citenamefont{Song et~al.}(2006)\citenamefont{Song, Lim, Jung,
  Santos, and Moodera}}]{Song2386}
\bibinfo{author}{\bibfnamefont{S.}~\bibnamefont{Song}},
  \bibinfo{author}{\bibfnamefont{S.}~\bibnamefont{Lim}},
  \bibinfo{author}{\bibfnamefont{M.}~\bibnamefont{Jung}},
  \bibinfo{author}{\bibfnamefont{T.}~\bibnamefont{Santos}}, \bibnamefont{and}
  \bibinfo{author}{\bibfnamefont{J.}~\bibnamefont{Moodera}},
  \bibinfo{journal}{J. Korean Phys. Soc.} \textbf{\bibinfo{volume}{49}},
  \bibinfo{pages}{2386} (\bibinfo{year}{2006}).

\bibitem[{\citenamefont{Sarma et~al.}(2003)\citenamefont{Sarma, Hwang, and
  Kaminski}}]{Das155201}
\bibinfo{author}{\bibfnamefont{S.~D.} \bibnamefont{Sarma}},
  \bibinfo{author}{\bibfnamefont{E.~H.} \bibnamefont{Hwang}}, \bibnamefont{and}
  \bibinfo{author}{\bibfnamefont{A.}~\bibnamefont{Kaminski}},
  \bibinfo{journal}{Phys. Rev. B} \textbf{\bibinfo{volume}{67}},
  \bibinfo{pages}{155201} (\bibinfo{year}{2003}).

\bibitem[{\citenamefont{Edmonds
  et~al.}(2004{\natexlab{a}})\citenamefont{Edmonds, Bogus{\l}awski, Wang,
  Campion, Novikov, Farley, Gallagher, Foxon, Sawicki, Dietl
  et~al.}}]{edmonds2004mn}
\bibinfo{author}{\bibfnamefont{K.}~\bibnamefont{Edmonds}},
  \bibinfo{author}{\bibfnamefont{P.}~\bibnamefont{Bogus{\l}awski}},
  \bibinfo{author}{\bibfnamefont{K.}~\bibnamefont{Wang}},
  \bibinfo{author}{\bibfnamefont{R.}~\bibnamefont{Campion}},
  \bibinfo{author}{\bibfnamefont{S.}~\bibnamefont{Novikov}},
  \bibinfo{author}{\bibfnamefont{N.}~\bibnamefont{Farley}},
  \bibinfo{author}{\bibfnamefont{B.}~\bibnamefont{Gallagher}},
  \bibinfo{author}{\bibfnamefont{C.}~\bibnamefont{Foxon}},
  \bibinfo{author}{\bibfnamefont{M.}~\bibnamefont{Sawicki}},
  \bibinfo{author}{\bibfnamefont{T.}~\bibnamefont{Dietl}},
  \bibnamefont{et~al.}, \bibinfo{journal}{Phys. Rev. Lett.}
  \textbf{\bibinfo{volume}{92}}, \bibinfo{pages}{037201}
  (\bibinfo{year}{2004}{\natexlab{a}}).

\bibitem[{\citenamefont{Lima et~al.}(2015)\citenamefont{Lima, Wahl, Augustyns,
  Silva, Costa, Houben, Edmonds, Gallagher, Campion, van Bael
  et~al.}}]{Lima012406}
\bibinfo{author}{\bibfnamefont{T.}~\bibnamefont{Lima}},
  \bibinfo{author}{\bibfnamefont{U.}~\bibnamefont{Wahl}},
  \bibinfo{author}{\bibfnamefont{V.}~\bibnamefont{Augustyns}},
  \bibinfo{author}{\bibfnamefont{D.}~\bibnamefont{Silva}},
  \bibinfo{author}{\bibfnamefont{A.}~\bibnamefont{Costa}},
  \bibinfo{author}{\bibfnamefont{K.}~\bibnamefont{Houben}},
  \bibinfo{author}{\bibfnamefont{K.}~\bibnamefont{Edmonds}},
  \bibinfo{author}{\bibfnamefont{B.}~\bibnamefont{Gallagher}},
  \bibinfo{author}{\bibfnamefont{R.}~\bibnamefont{Campion}},
  \bibinfo{author}{\bibfnamefont{M.}~\bibnamefont{van Bael}},
  \bibnamefont{et~al.}, \bibinfo{journal}{Appl. Phys. Lett.}
  \textbf{\bibinfo{volume}{106}}, \bibinfo{pages}{012406}
  (\bibinfo{year}{2015}).

\bibitem[{\citenamefont{Sawicki et~al.}(2010)\citenamefont{Sawicki, Chiba,
  Korbecka, Nishitani, Majewski, Matsukura, Dietl, and Ohno}}]{Sawichi2010}
\bibinfo{author}{\bibfnamefont{M.}~\bibnamefont{Sawicki}},
  \bibinfo{author}{\bibfnamefont{D.}~\bibnamefont{Chiba}},
  \bibinfo{author}{\bibfnamefont{A.}~\bibnamefont{Korbecka}},
  \bibinfo{author}{\bibfnamefont{Y.}~\bibnamefont{Nishitani}},
  \bibinfo{author}{\bibfnamefont{J.~A.} \bibnamefont{Majewski}},
  \bibinfo{author}{\bibfnamefont{F.}~\bibnamefont{Matsukura}},
  \bibinfo{author}{\bibfnamefont{T.}~\bibnamefont{Dietl}}, \bibnamefont{and}
  \bibinfo{author}{\bibfnamefont{H.}~\bibnamefont{Ohno}},
  \bibinfo{journal}{Nature Phys.} \textbf{\bibinfo{volume}{6}},
  \bibinfo{pages}{22} (\bibinfo{year}{2010}).

\bibitem[{\citenamefont{Fritzsche and Rothemund}(1975)}]{Fritzsche39}
\bibinfo{author}{\bibfnamefont{C.}~\bibnamefont{Fritzsche}} \bibnamefont{and}
  \bibinfo{author}{\bibfnamefont{W.}~\bibnamefont{Rothemund}},
  \bibinfo{journal}{Appl. Phys.} \textbf{\bibinfo{volume}{7}},
  \bibinfo{pages}{39} (\bibinfo{year}{1975}).

\bibitem[{\citenamefont{Scarpulla et~al.}(2008)\citenamefont{Scarpulla,
  Farshchi, Stone, Chopdekar, Yu, Suzuki, and Dubon}}]{Scarpulla073913}
\bibinfo{author}{\bibfnamefont{M.}~\bibnamefont{Scarpulla}},
  \bibinfo{author}{\bibfnamefont{R.}~\bibnamefont{Farshchi}},
  \bibinfo{author}{\bibfnamefont{P.}~\bibnamefont{Stone}},
  \bibinfo{author}{\bibfnamefont{R.}~\bibnamefont{Chopdekar}},
  \bibinfo{author}{\bibfnamefont{K.}~\bibnamefont{Yu}},
  \bibinfo{author}{\bibfnamefont{Y.}~\bibnamefont{Suzuki}}, \bibnamefont{and}
  \bibinfo{author}{\bibfnamefont{O.}~\bibnamefont{Dubon}}, \bibinfo{journal}{J.
  Appl. Phys.} \textbf{\bibinfo{volume}{103}}, \bibinfo{pages}{073913}
  (\bibinfo{year}{2008}).

\bibitem[{\citenamefont{Dietl et~al.}(2001)\citenamefont{Dietl, Ohno, and
  Matsukura}}]{dietl2001holePRB}
\bibinfo{author}{\bibfnamefont{T.}~\bibnamefont{Dietl}},
  \bibinfo{author}{\bibfnamefont{H.}~\bibnamefont{Ohno}}, \bibnamefont{and}
  \bibinfo{author}{\bibfnamefont{F.}~\bibnamefont{Matsukura}},
  \bibinfo{journal}{Phys. Rev. B} \textbf{\bibinfo{volume}{63}},
  \bibinfo{pages}{195205} (\bibinfo{year}{2001}).

\bibitem[{\citenamefont{Sawicki et~al.}(2004)\citenamefont{Sawicki, Matsukura,
  Idziaszek, Dietl, Schott, Ruester, Gould, Karczewski, Schmidt, and
  Molenkamp}}]{Sawicki245325}
\bibinfo{author}{\bibfnamefont{M.}~\bibnamefont{Sawicki}},
  \bibinfo{author}{\bibfnamefont{F.}~\bibnamefont{Matsukura}},
  \bibinfo{author}{\bibfnamefont{A.}~\bibnamefont{Idziaszek}},
  \bibinfo{author}{\bibfnamefont{T.}~\bibnamefont{Dietl}},
  \bibinfo{author}{\bibfnamefont{G.~M.} \bibnamefont{Schott}},
  \bibinfo{author}{\bibfnamefont{C.}~\bibnamefont{Ruester}},
  \bibinfo{author}{\bibfnamefont{C.}~\bibnamefont{Gould}},
  \bibinfo{author}{\bibfnamefont{G.}~\bibnamefont{Karczewski}},
  \bibinfo{author}{\bibfnamefont{G.}~\bibnamefont{Schmidt}}, \bibnamefont{and}
  \bibinfo{author}{\bibfnamefont{L.~W.} \bibnamefont{Molenkamp}},
  \bibinfo{journal}{Phys. Rev. B} \textbf{\bibinfo{volume}{70}},
  \bibinfo{pages}{245325} (\bibinfo{year}{2004}).

\bibitem[{\citenamefont{Thevenard et~al.}(2006)\citenamefont{Thevenard,
  Largeau, Mauguin, Patriarche, Lema\^{i}tre, Vernier, and
  Ferr\'{e}}}]{thevenard2006magnetic}
\bibinfo{author}{\bibfnamefont{L.}~\bibnamefont{Thevenard}},
  \bibinfo{author}{\bibfnamefont{L.}~\bibnamefont{Largeau}},
  \bibinfo{author}{\bibfnamefont{O.}~\bibnamefont{Mauguin}},
  \bibinfo{author}{\bibfnamefont{G.}~\bibnamefont{Patriarche}},
  \bibinfo{author}{\bibfnamefont{A.}~\bibnamefont{Lema\^{i}tre}},
  \bibinfo{author}{\bibfnamefont{N.}~\bibnamefont{Vernier}}, \bibnamefont{and}
  \bibinfo{author}{\bibfnamefont{J.}~\bibnamefont{Ferr\'{e}}},
  \bibinfo{journal}{Phys. Rev. B} \textbf{\bibinfo{volume}{73}},
  \bibinfo{pages}{195331} (\bibinfo{year}{2006}).

\bibitem[{\citenamefont{Stone et~al.}(2010)\citenamefont{Stone, Dreher, Beeman,
  Yu, Brandt, and Dubon}}]{stone2010interplay}
\bibinfo{author}{\bibfnamefont{P.}~\bibnamefont{Stone}},
  \bibinfo{author}{\bibfnamefont{L.}~\bibnamefont{Dreher}},
  \bibinfo{author}{\bibfnamefont{J.}~\bibnamefont{Beeman}},
  \bibinfo{author}{\bibfnamefont{K.}~\bibnamefont{Yu}},
  \bibinfo{author}{\bibfnamefont{M.}~\bibnamefont{Brandt}}, \bibnamefont{and}
  \bibinfo{author}{\bibfnamefont{O.}~\bibnamefont{Dubon}},
  \bibinfo{journal}{Phys. Rev. B} \textbf{\bibinfo{volume}{81}},
  \bibinfo{pages}{205210} (\bibinfo{year}{2010}).

\bibitem[{\citenamefont{Rushforth et~al.}(2008)\citenamefont{Rushforth, Wang,
  Farley, Campion, Edmonds, Staddon, Foxon, and Gallagher}}]{Rushforth073908}
\bibinfo{author}{\bibfnamefont{A.}~\bibnamefont{Rushforth}},
  \bibinfo{author}{\bibfnamefont{M.}~\bibnamefont{Wang}},
  \bibinfo{author}{\bibfnamefont{N.}~\bibnamefont{Farley}},
  \bibinfo{author}{\bibfnamefont{R.}~\bibnamefont{Campion}},
  \bibinfo{author}{\bibfnamefont{K.}~\bibnamefont{Edmonds}},
  \bibinfo{author}{\bibfnamefont{C.}~\bibnamefont{Staddon}},
  \bibinfo{author}{\bibfnamefont{C.}~\bibnamefont{Foxon}}, \bibnamefont{and}
  \bibinfo{author}{\bibfnamefont{B.}~\bibnamefont{Gallagher}},
  \bibinfo{journal}{J. Appl. Phys.} \textbf{\bibinfo{volume}{104}},
  \bibinfo{pages}{073908} (\bibinfo{year}{2008}).

\bibitem[{\citenamefont{Sawicki et~al.}(2005)\citenamefont{Sawicki, Wang,
  Edmonds, Campion, Staddon, Farley, Foxon, Papis, Kami{\'n}ska, Piotrowska
  et~al.}}]{sawicki2005plane}
\bibinfo{author}{\bibfnamefont{M.}~\bibnamefont{Sawicki}},
  \bibinfo{author}{\bibfnamefont{K.-Y.} \bibnamefont{Wang}},
  \bibinfo{author}{\bibfnamefont{K.}~\bibnamefont{Edmonds}},
  \bibinfo{author}{\bibfnamefont{R.}~\bibnamefont{Campion}},
  \bibinfo{author}{\bibfnamefont{C.}~\bibnamefont{Staddon}},
  \bibinfo{author}{\bibfnamefont{N.}~\bibnamefont{Farley}},
  \bibinfo{author}{\bibfnamefont{C.}~\bibnamefont{Foxon}},
  \bibinfo{author}{\bibfnamefont{E.}~\bibnamefont{Papis}},
  \bibinfo{author}{\bibfnamefont{E.}~\bibnamefont{Kami{\'n}ska}},
  \bibinfo{author}{\bibfnamefont{A.}~\bibnamefont{Piotrowska}},
  \bibnamefont{et~al.}, \bibinfo{journal}{Phys. Rev. B}
  \textbf{\bibinfo{volume}{71}}, \bibinfo{pages}{121302}
  (\bibinfo{year}{2005}).

\bibitem[{\citenamefont{Stone et~al.}(2008)\citenamefont{Stone, Bihler, Kraus,
  Scarpulla, Beeman, Yu, Brandt, and Dubon}}]{stone2008compensation}
\bibinfo{author}{\bibfnamefont{P.}~\bibnamefont{Stone}},
  \bibinfo{author}{\bibfnamefont{C.}~\bibnamefont{Bihler}},
  \bibinfo{author}{\bibfnamefont{M.}~\bibnamefont{Kraus}},
  \bibinfo{author}{\bibfnamefont{M.}~\bibnamefont{Scarpulla}},
  \bibinfo{author}{\bibfnamefont{J.}~\bibnamefont{Beeman}},
  \bibinfo{author}{\bibfnamefont{K.}~\bibnamefont{Yu}},
  \bibinfo{author}{\bibfnamefont{M.}~\bibnamefont{Brandt}}, \bibnamefont{and}
  \bibinfo{author}{\bibfnamefont{O.}~\bibnamefont{Dubon}},
  \bibinfo{journal}{Phys. Rev. B} \textbf{\bibinfo{volume}{78}},
  \bibinfo{pages}{214421} (\bibinfo{year}{2008}).

\bibitem[{\citenamefont{Glunk et~al.}(2009)\citenamefont{Glunk, Daeubler,
  Dreher, Schwaiger, Schoch, Sauer, Limmer, Brandlmaier, Goennenwein, Bihler
  et~al.}}]{glunk2009magnetic}
\bibinfo{author}{\bibfnamefont{M.}~\bibnamefont{Glunk}},
  \bibinfo{author}{\bibfnamefont{J.}~\bibnamefont{Daeubler}},
  \bibinfo{author}{\bibfnamefont{L.}~\bibnamefont{Dreher}},
  \bibinfo{author}{\bibfnamefont{S.}~\bibnamefont{Schwaiger}},
  \bibinfo{author}{\bibfnamefont{W.}~\bibnamefont{Schoch}},
  \bibinfo{author}{\bibfnamefont{R.}~\bibnamefont{Sauer}},
  \bibinfo{author}{\bibfnamefont{W.}~\bibnamefont{Limmer}},
  \bibinfo{author}{\bibfnamefont{A.}~\bibnamefont{Brandlmaier}},
  \bibinfo{author}{\bibfnamefont{S.}~\bibnamefont{Goennenwein}},
  \bibinfo{author}{\bibfnamefont{C.}~\bibnamefont{Bihler}},
  \bibnamefont{et~al.}, \bibinfo{journal}{Phys. Rev. B}
  \textbf{\bibinfo{volume}{79}}, \bibinfo{pages}{195206}
  (\bibinfo{year}{2009}).

\bibitem[{\citenamefont{Casiraghi et~al.}(2010)\citenamefont{Casiraghi,
  Rushforth, Wang, Farley, Wadley, Hall, Staddon, Edmonds, Campion, Foxon
  et~al.}}]{casiraghi2010tuning}
\bibinfo{author}{\bibfnamefont{A.}~\bibnamefont{Casiraghi}},
  \bibinfo{author}{\bibfnamefont{A.}~\bibnamefont{Rushforth}},
  \bibinfo{author}{\bibfnamefont{M.}~\bibnamefont{Wang}},
  \bibinfo{author}{\bibfnamefont{N.}~\bibnamefont{Farley}},
  \bibinfo{author}{\bibfnamefont{P.}~\bibnamefont{Wadley}},
  \bibinfo{author}{\bibfnamefont{J.}~\bibnamefont{Hall}},
  \bibinfo{author}{\bibfnamefont{C.}~\bibnamefont{Staddon}},
  \bibinfo{author}{\bibfnamefont{K.}~\bibnamefont{Edmonds}},
  \bibinfo{author}{\bibfnamefont{R.}~\bibnamefont{Campion}},
  \bibinfo{author}{\bibfnamefont{C.}~\bibnamefont{Foxon}},
  \bibnamefont{et~al.}, \bibinfo{journal}{Appl. Phys. Lett.}
  \textbf{\bibinfo{volume}{97}}, \bibinfo{pages}{122504}
  (\bibinfo{year}{2010}).

\bibitem[{\citenamefont{Edmonds
  et~al.}(2004{\natexlab{b}})\citenamefont{Edmonds, Farley, Campion, Foxon,
  Gallagher, Johal, van~der Laan, MacKenzie, Chapman, and
  Arenholz}}]{Edmonds4065}
\bibinfo{author}{\bibfnamefont{K.~W.} \bibnamefont{Edmonds}},
  \bibinfo{author}{\bibfnamefont{N.~R.~S.} \bibnamefont{Farley}},
  \bibinfo{author}{\bibfnamefont{R.~P.} \bibnamefont{Campion}},
  \bibinfo{author}{\bibfnamefont{C.~T.} \bibnamefont{Foxon}},
  \bibinfo{author}{\bibfnamefont{B.~L.} \bibnamefont{Gallagher}},
  \bibinfo{author}{\bibfnamefont{T.~K.} \bibnamefont{Johal}},
  \bibinfo{author}{\bibfnamefont{G.}~\bibnamefont{van~der Laan}},
  \bibinfo{author}{\bibfnamefont{M.}~\bibnamefont{MacKenzie}},
  \bibinfo{author}{\bibfnamefont{J.~N.} \bibnamefont{Chapman}},
  \bibnamefont{and} \bibinfo{author}{\bibfnamefont{E.}~\bibnamefont{Arenholz}},
  \bibinfo{journal}{Appl. Phys. Lett.} \textbf{\bibinfo{volume}{84}},
  \bibinfo{pages}{4065} (\bibinfo{year}{2004}{\natexlab{b}}).

\bibitem[{\citenamefont{Sch\"{u}tz et~al.}(1987)\citenamefont{Sch\"{u}tz,
  Wagner, Wilhelm, and Kienle}}]{Schutz737}
\bibinfo{author}{\bibfnamefont{G.}~\bibnamefont{Sch\"{u}tz}},
  \bibinfo{author}{\bibfnamefont{W.}~\bibnamefont{Wagner}},
  \bibinfo{author}{\bibfnamefont{W.}~\bibnamefont{Wilhelm}}, \bibnamefont{and}
  \bibinfo{author}{\bibfnamefont{P.}~\bibnamefont{Kienle}},
  \bibinfo{journal}{Phys. Rev. Lett.} \textbf{\bibinfo{volume}{58}},
  \bibinfo{pages}{737} (\bibinfo{year}{1987}).

\bibitem[{\citenamefont{Stone et~al.}(2006)\citenamefont{Stone, Scarpulla,
  Farshchi, Sharp, Haller, Dubon, Yu, Beeman, Arenholz, Denlinger
  et~al.}}]{Stone012504}
\bibinfo{author}{\bibfnamefont{P.~R.} \bibnamefont{Stone}},
  \bibinfo{author}{\bibfnamefont{M.~A.} \bibnamefont{Scarpulla}},
  \bibinfo{author}{\bibfnamefont{R.}~\bibnamefont{Farshchi}},
  \bibinfo{author}{\bibfnamefont{I.~D.} \bibnamefont{Sharp}},
  \bibinfo{author}{\bibfnamefont{E.~E.} \bibnamefont{Haller}},
  \bibinfo{author}{\bibfnamefont{O.~D.} \bibnamefont{Dubon}},
  \bibinfo{author}{\bibfnamefont{K.~M.} \bibnamefont{Yu}},
  \bibinfo{author}{\bibfnamefont{J.~W.} \bibnamefont{Beeman}},
  \bibinfo{author}{\bibfnamefont{E.}~\bibnamefont{Arenholz}},
  \bibinfo{author}{\bibfnamefont{J.~D.} \bibnamefont{Denlinger}},
  \bibnamefont{et~al.}, \bibinfo{journal}{Appl. Phys. Lett.}
  \textbf{\bibinfo{volume}{89}}, \bibinfo{pages}{012504}
  (\bibinfo{year}{2006}).

\bibitem[{\citenamefont{Zhou et~al.}(2012)\citenamefont{Zhou, Wang, Jiang,
  Weschke, and Helm}}]{Zhou093007}
\bibinfo{author}{\bibfnamefont{S.}~\bibnamefont{Zhou}},
  \bibinfo{author}{\bibfnamefont{Y.}~\bibnamefont{Wang}},
  \bibinfo{author}{\bibfnamefont{Z.}~\bibnamefont{Jiang}},
  \bibinfo{author}{\bibfnamefont{E.}~\bibnamefont{Weschke}}, \bibnamefont{and}
  \bibinfo{author}{\bibfnamefont{M.}~\bibnamefont{Helm}},
  \bibinfo{journal}{Appl. Phys. Express} \textbf{\bibinfo{volume}{5}},
  \bibinfo{pages}{093007} (\bibinfo{year}{2012}).

\bibitem[{\citenamefont{Thole et~al.}(1992)\citenamefont{Thole, Carra, Sette,
  and van~der Laan}}]{Thole1943}
\bibinfo{author}{\bibfnamefont{B.}~\bibnamefont{Thole}},
  \bibinfo{author}{\bibfnamefont{P.}~\bibnamefont{Carra}},
  \bibinfo{author}{\bibfnamefont{F.}~\bibnamefont{Sette}}, \bibnamefont{and}
  \bibinfo{author}{\bibfnamefont{G.}~\bibnamefont{van~der Laan}},
  \bibinfo{journal}{Phys. Rev. Lett.} \textbf{\bibinfo{volume}{68}},
  \bibinfo{pages}{1943} (\bibinfo{year}{1992}).

\bibitem[{\citenamefont{O'Brien and Tonner}(1994)}]{Obrien12672}
\bibinfo{author}{\bibfnamefont{W.~L.} \bibnamefont{O'Brien}} \bibnamefont{and}
  \bibinfo{author}{\bibfnamefont{B.~P.} \bibnamefont{Tonner}},
  \bibinfo{journal}{Phys. Rev. B} \textbf{\bibinfo{volume}{50}},
  \bibinfo{pages}{12672} (\bibinfo{year}{1994}).

\bibitem[{\citenamefont{Matsukura et~al.}(1998)\citenamefont{Matsukura, Ohno,
  Shen, and Sugawara}}]{MatsukuraR2037}
\bibinfo{author}{\bibfnamefont{F.}~\bibnamefont{Matsukura}},
  \bibinfo{author}{\bibfnamefont{H.}~\bibnamefont{Ohno}},
  \bibinfo{author}{\bibfnamefont{A.}~\bibnamefont{Shen}}, \bibnamefont{and}
  \bibinfo{author}{\bibfnamefont{Y.}~\bibnamefont{Sugawara}},
  \bibinfo{journal}{Phys. Rev. B} \textbf{\bibinfo{volume}{57}},
  \bibinfo{pages}{R2037} (\bibinfo{year}{1998}).

\bibitem[{\citenamefont{Kaminski and Sarma}(2003)}]{Kaminski235210}
\bibinfo{author}{\bibfnamefont{A.}~\bibnamefont{Kaminski}} \bibnamefont{and}
  \bibinfo{author}{\bibfnamefont{S.~D.} \bibnamefont{Sarma}},
  \bibinfo{journal}{Phys. Rev. B} \textbf{\bibinfo{volume}{68}},
  \bibinfo{pages}{235210} (\bibinfo{year}{2003}).

\bibitem[{\citenamefont{Esch et~al.}(1997)\citenamefont{Esch, Bockstal, Boeck,
  Verbanck, van Steenbergen, Wellmann, Grietens, Bogaerts, Herlach, and
  Borghs}}]{Esch13103}
\bibinfo{author}{\bibfnamefont{A.~V.} \bibnamefont{Esch}},
  \bibinfo{author}{\bibfnamefont{L.~V.} \bibnamefont{Bockstal}},
  \bibinfo{author}{\bibfnamefont{J.~D.} \bibnamefont{Boeck}},
  \bibinfo{author}{\bibfnamefont{G.}~\bibnamefont{Verbanck}},
  \bibinfo{author}{\bibfnamefont{A.}~\bibnamefont{van Steenbergen}},
  \bibinfo{author}{\bibfnamefont{P.}~\bibnamefont{Wellmann}},
  \bibinfo{author}{\bibfnamefont{B.}~\bibnamefont{Grietens}},
  \bibinfo{author}{\bibfnamefont{R.}~\bibnamefont{Bogaerts}},
  \bibinfo{author}{\bibfnamefont{F.}~\bibnamefont{Herlach}}, \bibnamefont{and}
  \bibinfo{author}{\bibfnamefont{G.}~\bibnamefont{Borghs}},
  \bibinfo{journal}{Phys. Rev. B} \textbf{\bibinfo{volume}{56}},
  \bibinfo{pages}{13103} (\bibinfo{year}{1997}).

\end{thebibliography}

\end{document}